\documentclass{cambridge6A}
  \usepackage{graphicx}
  
  \usepackage{euscript}
\usepackage{amssymb}
\usepackage{amsfonts}
\usepackage{amsbsy}
\usepackage{amsmath}
\usepackage{epsfig}
\usepackage{amsthm}
\usepackage{amscd}
\usepackage{amstext}
\usepackage{verbatim}

\def\ben{\begin{equation}}
\def\een{\end{equation}}

\def\qtr{{\textstyle{1\over4}}}
\let\a=\alpha \let\b=\beta \let\g=\gamma \let\d=\delta 
   \let\k=\kappa
     
\let\s=\sigma

\let\w=\omega \let\G=\Gamma

\let\pa=\partial
\def\be{\begin{equation}}
\def\ee{\end{equation}}
\def\beq{\begin{equation}}
\def\eeq{\end{equation}}
\def\ba{\begin{array}}
\def\ea{\end{array}}

\def\dalemb#1#2{{\vbox{\hrule height .#2pt
       \hbox{\vrule width.#2pt height#1pt \kern#1pt
               \vrule width.#2pt}
       \hrule height.#2pt}}}

\newcommand{\bea}{\begin{eqnarray}}
\newcommand{\eea}{\end{eqnarray}}
\newcommand{\tr}{{\rm tr} }

\renewcommand*\thesection{\arabic{section}}
\renewcommand*\thesubsection{\arabic{section}.\arabic{subsection}}
\numberwithin{equation}{section}
\renewcommand{\thefigure}{\arabic{figure}}  
  
\begin{document}

\chapterauthor{Sean A. Hartnoll \\ ~ \\
\textit{Center for the Fundamental Laws of Nature, Harvard University,\\
Cambridge, MA 02138, USA}}

\chapter*{Horizons, holography and condensed matter}

\contributor{Sean A. Hartnoll
\affiliation{Center for the Fundamental Laws of Nature, Harvard University,\\
Cambridge, MA 02138, USA}
}

\begin{abstract}\small
The holographic correspondence creates an interface between classical gravitational physics and the dynamics
of strongly interacting quantum field theories. This chapter will relate the physics of charged, asymptotically Anti-de Sitter
spacetimes to the phenomenology of low temperature critical phases of condensed matter. Common essential features will
characterise both the gravitational and field theoretic systems. Firstly, an emergent scaling symmetry at the lowest energy scales
appears as an emergent isometry in the interior, `near horizon' regime of the spacetime. Secondly, the field theoretic distinction
between fractionalized and mesonic phases appears as the presence or absence of a charge-carrying horizon in the spacetime.
A perspective grounded in these two characteristics allows a unified presentation of `holographic superconductors', `electron stars' and
`charged dilatonic spacetimes'.
\end{abstract}
 
\copyrightline{Chapter of the book \textit{Black Holes in Higher Dimensions} to
be published by Cambridge University Press (editor: G. Horowitz)} 
 
\section{Introduction}

Consistent theories of quantum gravity in spacetimes that asymptote to Anti-de Sitter ($AdS$) spacetime are equivalent to 
quantum field theories defined on the conformal boundary of the spacetime \cite{Maldacena:1997re}. A pedagogical
discussion of this `holographic correspondence' may be found in \cite{Polchinski:2010hw, McGreevy:2009xe} and in the chapter in this volume. While some of the deeper questions arising from the correspondence remain to be understood from first principles,
the conceptual `Gestalt switch' involved in viewing physical processes simultaneously from a gravitational and a field theoretic
perspective has provided an invaluable source of physical intuition as well as computational power. In particular, in a `large $N$' limit of quantum field theories, to be recalled shortly, the gravitational description becomes weakly curved and the tools of general relativity may be harnessed.

This chapter will be concerned with black holes in four dimensional asymptotically $AdS$ spacetimes. By focussing on charged, planar black holes, we will establish an interface with a rich phenomenology of 2+1 dimensional quantum field theories that has been widely studied in condensed matter physics. Planarity of the horizon will translate into the statement that the dual quantum field theory propagates on a background Minkowski spacetime in 2+1 dimensions. The perhaps more familiar spherical foliation of asymptotically $AdS$ spacetimes would have corresponded to considering quantum field theories on a spatial sphere. This complicates the field theoretic physics by introducing a scale, the radius of the sphere, and also does not correspond to a situation of significant interest in condensed matter physics at present. The charge of the black hole will translate into the fact that the field theory is in a state with a nonzero charge density. This charge density is to be thought of as the `stuff' of condensed matter physics; a proxy for, for instance, the fluid of electrons in a metal.

In this chapter we will pursue a gravitational approach to the following fundamental field theoretic question: Consider a general quantum field theory in a state with a finite charge density. How might one attempt to classify all possible gapless low temperature phases of matter that can arise? Free field theories exhibit two well known low temperature phases. A free charged boson will undergo Bose-Einstein condensation, spontaneously breaking the charge symmetry. The gapless degree of freedom is consequently a Goldstone boson. Free charged fermions, in contrast, cannot macroscopically occupy their ground state, but rather build up a Fermi surface. The gapless degrees of freedom are then particle-hole excitations of the Fermi surface. The dynamics of Goldstone bosons and Fermi surface excitations is tightly constrained by kinematics and well understood. Beyond free or weakly interacting theories, however, the question of possible phases of matter becomes more difficult -- this is where a gravitational perspective may make itself useful.

After a brief review of the holographic correspondence and of some challenges and expectations from condensed matter theory, 
we will translate our guiding question into gravitational terms. Take a gravitational theory with some specific matter content that admits asymptotically $AdS$ solutions. Require the spacetime to have a net asymptotic electric flux, i.e. require the spacetime to be charged.
What, then, is the thermodynamically dominant spacetime that sources this charge? The basic dichotomy that we will discuss is whether the electric flux emanates from behind a charged horizon in the interior of the spacetime or whether it is explicitly sourced by charged matter.
The gravitational physics resolving this tension will be that of charged superradiance instabilities. Returning to a field theory perspective at the end of the chapter, we will argue that this distinction is the gravitational realization of the field theoretic distinction between `fractionalised'  phases versus `mesonic' phases. The precise meaning of these terms will be made clear in what follows.

\section{Preliminary holographic notions}
\label{sec:prelim}

There exist certain quantum field theories in which the locality of the renormalisation group (RG) flow can be (usefully) geometrically realised.
This is a feature of the holographic correspondence that will be central to our discussion. The basic idea is to append an extra
spatial dimension to the spacetime of the quantum field theory. This extra dimension will correspond to the RG scale as illustrated in figure \ref{fig:rga} below.
\begin{figure}[h!]
\begin{center}
\includegraphics[scale=.3]{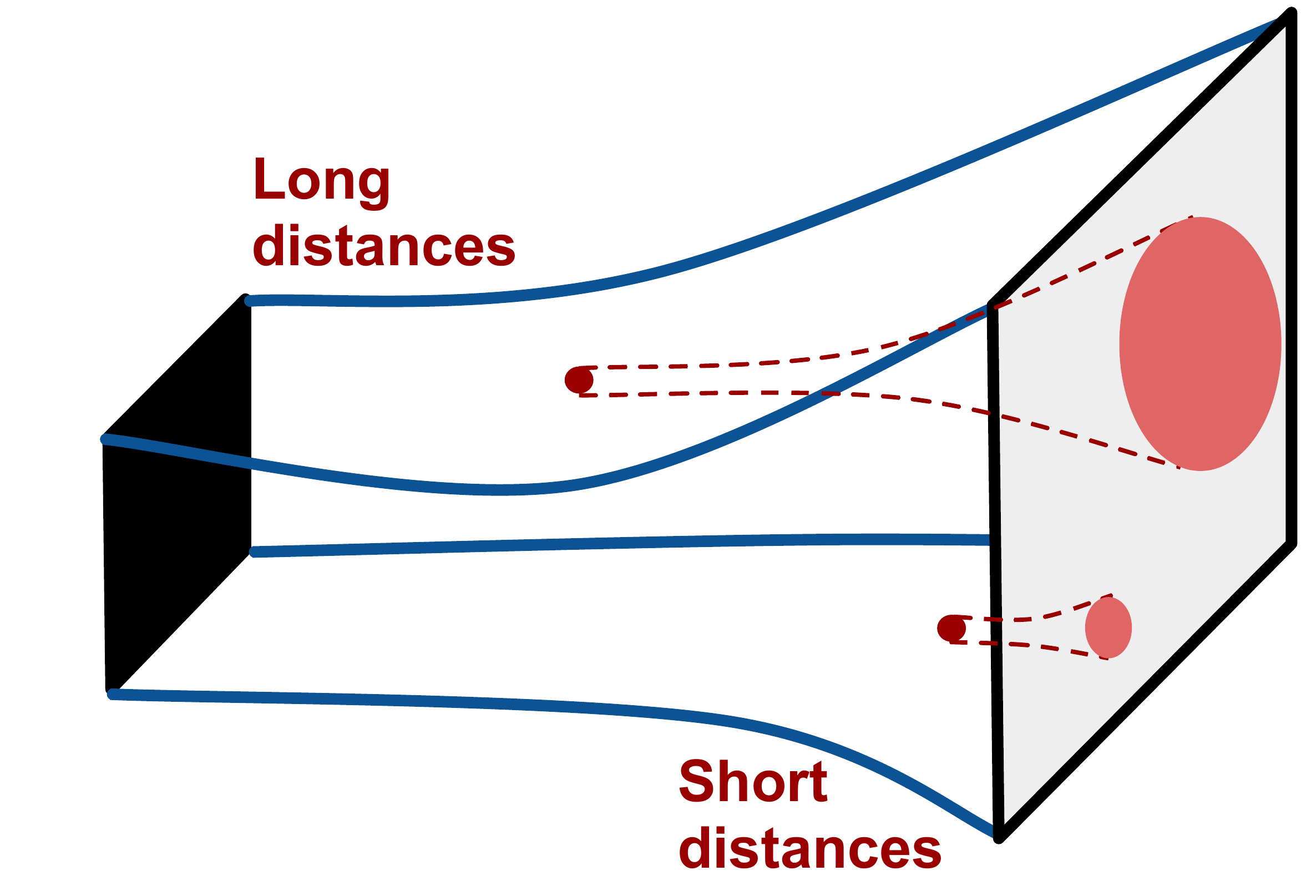}
\end{center}
\vspace{-.5cm}
\caption{
The extra radial dimension in holography corresponds to the renormalisation group scale. Processes in the interior determine long distance physics, the IR, of the dual field theory while processes near the boundary control the short distance, or UV, physics.\label{fig:rga}}
\end{figure}
In contrast to the fixed `boundary' field theory spacetime, the `bulk' spacetime with an extra dimension will be dynamical. The boundary conditions set at infinity in the bulk correspond to the UV values of couplings in the field theory. Solving the gravitational equations of motion is dual to following the RG flow down in energy scales. A modern presentation of the holographic renormalisation group may be found in \cite{Heemskerk:2010hk, Faulkner:2010jy}. For our purposes we will only need the mental picture of figure \ref{fig:rga} as a way of organising our thoughts about asymptotically $AdS$ spacetimes. The asymptotic spacetime describes the UV of the quantum field theory while the interior of the spacetime describes the IR.

At this point we can understand why $AdS$ spacetime plays a privileged role in discussions of holography. The simplest quantum field theories are those that exhibit no RG flow at all, i.e. that are scale invariant. $AdS$ spacetime is the geometrisation of this invariance for a relativistic quantum field theory. In our planar coordinates, $AdS$ spacetime takes the form
\be\label{eq:ads}
ds^2 = \frac{L^2}{r^2} \left(- dt^2 + dr^2 + dx^2 + dy^2 \right) \,.
\ee
This spacetime is invariant under
\be\label{eq:adsscaling}
r \to \lambda r \,, \quad \{t,x,y\} \to \lambda \{t, x,y\} \,.
\ee
Therefore if we follow an RG flow in the boundary theory, by rescaling the field theory coordinates $\{t, x,y\}$, we must simultaneously move into the bulk.\footnote{In the coordinates of (\ref{eq:ads}), and throughout this chapter, the conformal boundary of spacetime is at $r \to 0$, and $r$ increases as one moves into the space. This is essentially the inverse of the radial coordinate used elsewhere in this book. Such a coordinate has been more conventionally denoted by $z$. We need, however, to keep $z$ free to denote the dynamical critical exponent below. We note that the coordinate $r$ has units of length.} When we do this, the spacetime does not change. More generally, spacetimes will not be scale invariant but rather only asymptotically $AdS$, so that the dual field theory approaches a fixed point at high energies.

One must ask exactly which class of quantum field theories admit a holographic description, with the renormalisation group classically geometrised. Insofar as the answer to this question is known, two properties are key. Firstly, the theory must admit a large $N$ expansion. Secondly, in this large $N$ limit `most' of the operators in the theory must acquire parametrically large anomalous dimensions \cite{Heemskerk:2009pn}. The role of the large $N$ limit is that it implies an underlying `master' classical field configuration that dominates the path integral \cite{ed}. The expectation values of operators must factorise into the products of the expectation values of `single trace' operators, the effective classical fields, to leading order at large $N$. These single trace operators will correspond to classical single particle states in the bulk description. In general, one must still deal with infinitely many such classical fields, as occurs for example with the large $N$ limit of the $O(N)$ model \cite{Klebanov:2002ja}. The additional condition that all except a small handful of the single trace operators acquire parametrically large anomalous dimensions will translate into the bulk statement that all except a handful of the classical bulk fields become parametrically heavy and may therefore be ignored for many questions. It is clear that this additional fact requires the large $N$ field theory to be strongly interacting. A finite number of classical fields in the bulk can then be described by a local classical action, whose `radial' local equations of motion have a chance of realising the local renormalisation group flow equations for finitely many single trace operators and the multitrace operators they generate \cite{Heemskerk:2009pn,Heemskerk:2010hk, Faulkner:2010jy}.

Many theories are known for which the two properties of the previous paragraph hold true. Some, such as ${\mathcal{N}}=4$ super Yang-Mills theory in $3+1$ dimensions \cite{Maldacena:1997re}, the `ABJM' class of ${\mathcal{N}}=6$ gauge theories in $2+1$ dimensions \cite{Aharony:2008ug}, and their many cousins with less supersymmetry, have explicitly known Lagrangian descriptions. The schematic form of the Lagrangians is
\be\label{eq:N4}
{\mathcal{L}}\sim \tr \left(F^2 + \left(\partial \Phi \right)^2 + i \bar \Psi \Gamma \cdot \partial \Psi + g^2 [\Phi,\Phi]^2 + i g \bar \Psi [\Phi, \Psi] \right) \,.
\ee
This schematic expression is attempting to convey the following features: The theory has a large $N$ nonabelian gauge group with field strength $F$. The theory contains adjoint bosonic ($\Phi$) and fermionic ($\Psi$) matter. The matter will typically come in multiple flavors and with various patterns of interactions. The matter fields can be charged under global symmetries. Throughout this chapter we will assume the existence of a global $U(1)$ symmetry in the theory. We will add a chemical potential for this symmetry and thereby induce a charge density.

For the vast majority of quantum field theories with classical gravity duals, however, we do not know the field theory Lagrangian explicitly. Any consistent theory of quantum gravity with a superselection sector described by asymptotically $AdS$ spacetimes will define a dual quantum field theory. Very many such constructions are believed to exist, these form the `landscape' of string vacua, e.g. \cite{Denef:2008wq}. 
Rather than via a Lagrangian, the dual field theories are characterised by their spectrum of operators and the correlation functions of these operators. One can cogently argue that such a description of a theory, in terms of operators and correlators, is better suited to strongly interacting theories than a Lagrangian description, which describes the physics in terms of somewhat fictitious weakly interacting fields.

As this chapter is a gravitational perspective on holographic physics, we shall look at general features of charged asymptotically $AdS$ spacetimes without concerning ourselves with the specific dual field theories involved. Some of the results we will discuss have been embedded into `actually existing' concrete holographic dualities, while others remain to be so realised.

We have stated that the large $N$ field theory limit is a classical limit and that the bulk gravitational description provides the anticipated classical description. Let us briefly see how this works out more explicitly. The simplest theory that has the $AdS$ metric (\ref{eq:ads}) as solution is Einstein gravity with a negative cosmological constant
\be
{\mathcal L} = \frac{1}{2 \k^2} \left(R + \frac{6}{L^2} \right) \,.
\ee
To exhibit the large number of degrees of freedom of the large $N$ limit in a universal way, we can heat up the field theory. This is achieved by considering a (planar) black hole in the interior of $AdS$ spacetime. The Schwarzschild-AdS solution is 
\be\label{eq:adsBH}
ds^2 = \frac{L^2}{r^2} \left(- f(r) dt^2 + \frac{dr^2}{f(r)} + dx^2 + dy^2 \right) \,.
\ee
Here the metric function and corresponding field theory temperature are
\be\label{eq:fT}
f(r) =1 - \left( \frac{r}{r_+} \right)^{3} \,, \qquad T = \frac{3}{4 \pi r_+} \,.
\ee
As usual the temperature can be determined as the inverse period of the Euclidean time circle that renders the Euclidean Schwarzschild-AdS solution regular at the horizon $r=r_+$. A nonzero temperature is an IR phenomenon. Consequently we see in (\ref{eq:fT}) that the UV metric ($r\to 0$) is not altered by the presence of a horizon, while the IR interior of the spacetime is qualitatively changed. From the Euclidean solution we can compute the free energy of the field theory by evaluating the gravitational path integral on the Euclidean saddle point, so that $Z = e^{-S_E}$,
\be
F = - T \log Z = T S_E = - \frac{(4 \pi)^3}{2 \cdot 3^3} \frac{L^2}{\k^2} V_2 T^3 \,.
\ee
In evaluating the on shell action one must `holographically renormalise' volume divergences (see e.g. \cite{Hartnoll:2009sz}). In the above expression $V_2$ is the spatial volume of the field theory. The temperature dependence is determined by dimensional analysis -- we started with a scale invariant theory and so the temperature is the only scale. The coefficient of the temperature scaling gives a measure of the number of degrees of freedom of the theory. In the large $N$ limit we therefore expect
\be\label{eq:central}
\frac{L^2}{\k^2} \gg 1 \,.
\ee
And indeed, this is the classical gravitational limit in which the $AdS$ curvatures are small in Planck units.

\section{Brief condensed matter motivation}
\label{sec:motivation}

This section will introduce the quantum field theory problem of a nonzero density of fermions coupled to a gapless boson.
We will firstly explain why such a system is of interest in condensed matter physics and secondly why it is a challenging system
to study in 2+1 spacetime dimensions.

Consider the conduction electrons in a metal. At the lattice energy scale we are faced with a strongly interacting many body problem.
Landau argued that the solution to this many body problem would simplify for physics at the lowest energy scales, see e.g. \cite{anderson}.
One step in Landau's argument followed from the Pauli exclusion principle obeyed by the density of fermions. The conduction electrons would be forced to build up a Fermi surface in momentum space, and consequently the lowest energy fermionic excitations would correspond to removing or adding a fermion to the top of the Fermi sea. These excitations would not live at the origin of momentum space but rather at the Fermi momentum $k = k_F$. The reduced phase space available to fermions to scatter close to the Fermi surface then results in a suppression of the effects of interactions. In a modern renormalisation group language, the `Fermi liquid theory' of the low energy excitations of fermions about a Fermi surface turns out to be an IR free fixed point, independently of the strength of electron interactions at the UV lattice scale \cite{Polchinski:1992ed, shankar}. It therefore provides a robust weakly interacting starting point from which physics such as, for instance, superconducting pairing instabilities may be studied.

In recent years, a large number of experiments on many different families of materials, including several families of nonconventional superconductors, have indicated that in many situations of significant interest, Fermi liquid theory does not adequately describe the low energy electronic physics. One famous indication of this fact is that the observed electrical resistivity is larger than in Fermi liquid theory, often scaling like $T$ rather than the predicted $T^2$ at low temperatures.\footnote{For completeness we should note that to obtain a nonzero resistivity, translational invariance must be broken by adding e.g. contact `umklapp' interactions to the Fermi liquid theory.} See e.g. \cite{criticality} for an overview of such measurements. This may suggest the presence of additional low energy excitations capable of efficiently scattering the low energy current-carying fermions. As we have just recalled that fermions are unable to scatter efficiently near the Fermi surface, one is led to consider the existence of additional gapless bosonic degrees of freedom, taking the system outside the low energy universality class of Fermi liquid theory.

Bosonic excitations can arise as collective modes of the UV electrons. However, in order for the bosons to be gapless, the system must either be tuned to a `quantum critical point' at which the mass of the boson vanishes, or else there must be a kinematical constraint leading to a `critical phase' where the boson can remain massless over a range of parameter space. A quantum critical point separates different zero temperature phases of matter. If one of the phases is characterised by an order parameter, then at the critical point, fluctuations of the order parameter will become massless, and these are the modes that can scatter the fermions \cite{review, subirbook}. Critical phases are perhaps more closely related to the types of theories with holographic duals, and so we will discuss them in a little more detail. They will also connect directly to our later discussion of spacetimes with and without charged event horizons.

A natural way to describe the more robust gapless bosons of critical phases is as deconfined gauge fields, whose masslessness is protected by gauge invariance. Gauge fields can emerge as collective excitations of electrons when the microscopic lattice theory contains constraints, such as e.g. forbidding double occupancy of lattice sites \cite{mottreview}. In terms of the creation operator $c^\dagger_{i\s}$ for an election with spin $\sigma = \uparrow, \downarrow$, the no double occupancy constraint at each site $i$ reads
\be\label{eq:cons1}
\sum_\sigma c^\dagger_{i\s} c_{i\s} \leq 1 \,.
\ee
Such constraints can be elegantly recast as equalities rather than inequalities using a redundant mathematical description of the electron as a composite of a spinon $f_{i\s}^\dagger$ and holon $b_i^\dagger$ particle. The constraint (\ref{eq:cons1}) becomes the statement that at each site there must either be an up spin, a down spin, or a hole (i.e. no electrons). Let us write this as
\be\label{eq:gauss}
\sum_\sigma f^\dagger_{i\s} f_{i\s} + b^\dagger_{i} b_{i} = 1\,.
\ee
This is equivalent to `fractionalizing' the electron into its spin and charge degrees of freedom by writing
\be\label{eq:cfb}
c_{i\s} = f_{i\s} b^\dagger_{i} \,.
\ee
This description is redundant because the local (in time and space) transformation
\be\label{eq:sym}
f_{i\s}(t) \to e^{i \theta(t)} f_{i\s}(t) \,, \qquad b_{i}(t) \to e^{i \theta(t)} b_{i}(t) \,,
\ee
leaves the physical field $c_{i\s}$ invariant. This redundancy must be cancelled out of the theory by gauging the symmetry (\ref{eq:sym}).
Upon gauging, the constraint (\ref{eq:gauss}) appears as the local Gauss law for the total gauge charge at each site.

The main purposes of the previous paragraph were firstly to explain how local microscopic constraints motivate the emergence of gauge fields, and secondly to introduce the notion of fractionalization, in which a gauge-invariant fermion, $c$, is expressed as a composite of a gauge-charged fermion, $f$, and a gauge-charged boson, $b$. A description of the system in terms of an emergent gauge boson and gauge charged bosons and fermions starts to take us close to the class of theories discussed above around (\ref{eq:N4}). The substantial difference between the continuum limit of the theory discussed here and that of (\ref{eq:N4}) is that we are discussing an emergent $U(1)$ gauge field, while the holographic theories typically have $SU(N)$ gauge fields, with $N$ large. Such will be the price of theoretical control over computations.

Suppose we are granted a critical phase described by an emergent photon. A simple low energy effective theory describing the interaction of this photon with the fermionic excitations of a Fermi surface is QED at finite chemical potential $\mu$
\be\label{eq:QED}
{\mathcal{L}} = - \frac{1}{4} F^2 + \bar \psi \left(\gamma \cdot (i \pa + A) + \g^0 \mu \right) \psi \,.
\ee
While 2+1 dimensional Maxwell theories tend to confine \cite{Polyakov:1976fu}, it is believed that the presence of a Fermi surface dynamically suppresses the instantons responsible for confinement, e.g. \cite{sungsik0, Unsal:2008sc}. Therefore the theory (\ref{eq:QED}) is an example of a 2+1 dimensional theory that, at energy scales well below the chemical potential scale, should describe excitations of a Fermi surface interacting with a gapless boson.

A useful recent discussion of the theory (\ref{eq:QED}) at the lowest energy scales, including references to an extensive earlier literature, can be found in \cite{sungsik}. It had been know for some time that the theory flowed to strong coupling, but it was believed that the RG flow could be reigned in by means of a large $N$ expansion in which the number of fermion fields was taken large (this is quite different from the 't Hooft matrix large $N$ that is being discussed in the rest of this chapter). An important result of \cite{sungsik} was that this large $N$ expansion broke down at high loop order, due to particular kinematic effects associated with the presence of a Fermi surface. This has the consequence that controlling the low energy physics seems to require directly confronting strong interactions. It was subsequently observed that such effects occur also in other models of bosons coupled to 2+1 dimensional Fermi surfaces, where the boson described gapless order parameter fluctuations rather than an emergent gauge field \cite{Metlitski:2010pd, Metlitski:2010vm}. Obtaining a controlled description of the low energy dynamics of a finite density of fermions coupled to a gapless boson remains at present a formidable quantum field theoretic problem with potentially important consequences for exotic states of matter. Attempts to address this problem using more conventional field theoretical frameworks than holography include \cite{mross}.

\begin{figure}[h!]
\begin{center}
\includegraphics[scale=.5]{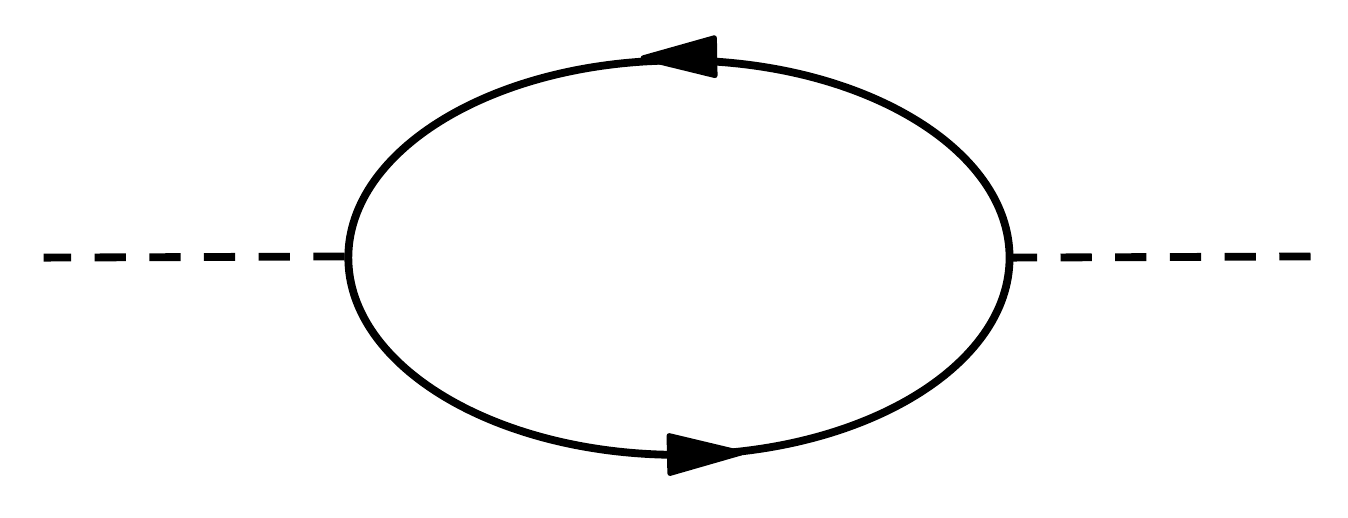}
\end{center}
\vspace{-.5cm}
\caption{One loop correction to the boson propagator from fermions.\label{fig:loop}}
\end{figure}

While field theoretic computations in the model (\ref{eq:QED}) are not controlled at low energies, it is possible to gain qualitative insight from uncontrolled perturbative computations. A basic quantity to consider is the propagator for the bosonic field. In the QED example (\ref{eq:QED}) this is the transverse gauge field $\vec A$ (the temporal component becomes gapped due to screening by the density of fermions). In general it is best to consider gauge-invariant quantities, but this correlator has the same form in theories at quantum critical points in which the gapless boson is gauge invariant. Classically, the inverse boson propagator has the schematic form $D(\w,k)^{-1} = \w^2 + k^2$. The leading correction is given by the loop of fermions shown in figure \ref{fig:loop}.
Evaluating this self energy diagram, the inverse propagator to leading order at low (Euclidean) energies $\w$ and momenta $k$ becomes
\be\label{eq:Dz}
D(\w,k)^{-1} =  \g \frac{|\w|}{|k|} + k^2 \,.
\ee
The rather non-analytic structure of the low energy propagator is possible because the chemical potential $\mu$ breaks Lorentz invariance. The main point we wish to take home is that while the UV propagator has the Lorentzian scale invariance $\{t,|x|\} \to \lambda \{t,|x|\}$, the emergent IR scaling of the propagator (\ref{eq:Dz}) is
\be\label{eq:z3}
t \to \lambda^3 t \,, \qquad |x| \to \lambda |x| \,.
\ee
We are simplifying a little here, the correct scaling is locally anisotropic in momentum due to the presence of a Fermi surface; the interested reader is referred to \cite{sungsik}. The scaling in (\ref{eq:z3}) is said to correspond to a dynamical critical exponent $z=3$. In general, $z$ denotes the relative scaling of space and time. The one loop result $z=3$ is not typically protected from order one corrections at higher loop order. The phenomenon whereby the interaction of the boson with a density of fermions causes a strong frequency dependence in the boson propagator is called Landau damping.

To summarize this brief motivation from condensed matter physics: it is of interest to understand the behaviour of gapless bosons coupled to the excitations of a Fermi surface. One example of this occurs when a metallic system exhibits an emergent gauge symmetry. Quantum field theories describing such bosons and fermions are typically strongly interacting at low energies in 2+1 dimensions and resilient to conventional field theoretic techniques. An important quantity characterizing the emergent strongly interacting low energy theory is the dynamical critical exponent $z$.

\section{Holography with a chemical potential}
\label{sec:holo}

The upshot of the previous section is that we would like to have a controlled framework to study the low energy physics of gapless bosons coupled to a density of fermions. We have seen that field theories with holographic gravity duals typically have the ingredients necessary to approach this question: $SU(N)$ gauge fields that are massless and matter charged under a global symmetry that can be placed at a chemical potential to induce a charge density. In this section we will describe how to set up the gravitational version of this field theoretic problem.

We wish to consider finite density states of matter. This means that $\langle J^t \rangle \neq 0$, where $J^t$ is the charge density operator for a global $U(1)$ symmetry in the field theory. A finite charge density is induced by holding the system at a nonzero chemical potential. Recall that this means we add to the Lagrangian of the theory the term $\Delta {\mathcal L} = \mu J^t$. The most basic explicit entry in the holographic dictionary is that single trace operators in the quantum field theory are in one to one correspondence with fields in the gravitating bulk. In particular, see e.g. \cite{Hartnoll:2009sz}, the charge density operator $J^t$ is dual to the time component of a Maxwell field $A_t$ in the bulk. The connection between bulk and field theory quantities works as follows: In our classical limit, the gauge potential will obey Maxwell's equations in the bulk. For a given solution to the bulk equations of motion, the corresponding values of $\mu$ and $\langle J^t \rangle$ are read off from the asymptotic near-boundary ($r \to 0$) behaviour of the solution as
\be
A_t(r) = \mu + \langle J^t \rangle r + \cdots \,. 
\ee
Thus the chemical potential and charge density are the boundary values of the Maxwell potential and electric flux, respectively
\be
\mu = \lim_{r \to 0} A_t \,, \qquad \langle J^t \rangle = \lim_{r \to 0} F_{rt} \,.
\ee
Therefore to impose that the quantum field theory is at nonzero density, we must impose that the dual spacetime has an electric flux at infinity.

\begin{figure}[h!]
\begin{center}
\includegraphics[scale=.3]{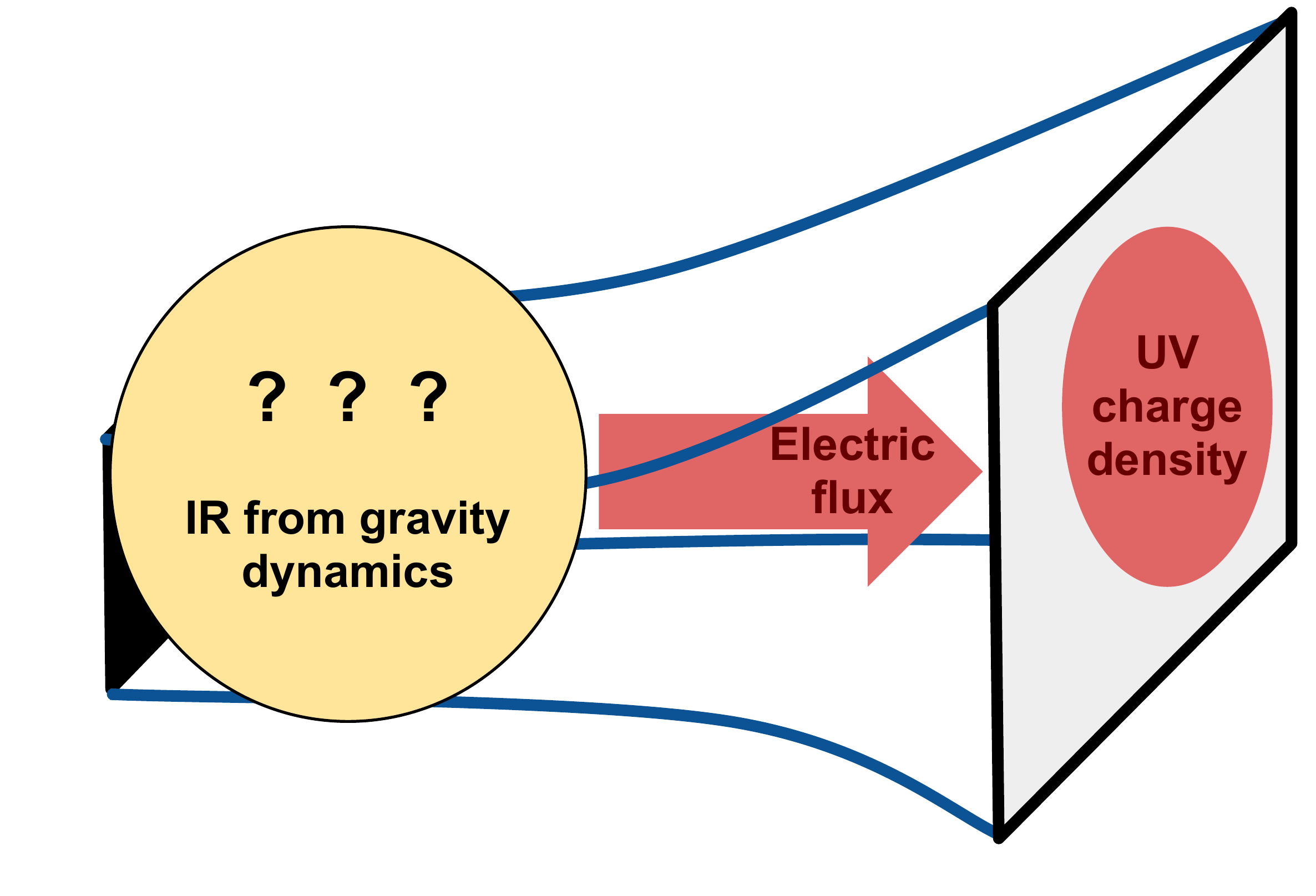}
\end{center}
\vspace{-.5cm}
\caption{
The basic question in finite density holography: use the gravitational equations to motion to find the interior IR geometry given the boundary condition that there is an electric flux at infinity.\label{fig:flux}}
\end{figure}

As explained in section \ref{sec:prelim} above, the asymptotic boundary conditions on the spacetime correspond to the UV starting point of the field theoretic RG flow. Integrating the bulk equations of motion into the interior of the spacetime corresponds to flowing down to the low energy IR of the theory. This is the regime where traditional field theoretic approaches run into difficulties and so we would like to understand what  holography has to say. The holographic framework is illustrated in figure \ref{fig:flux} and will guide the remainder of this chapter. We will find that the question of how to `fill in' the spacetime, given electric flux at the boundary, leads to gravitational physics that is interesting in its own right.

\section{The planar Reissner-Nordstr\"om-AdS black hole}

The minimal framework capable of describing the physics of electric flux in an asymptotically $AdS$ geometry is Einstein-Maxwell theory with a negative cosmological constant \cite{Chamblin:1999tk}. The Lagrangian density can be written
\be\label{eq:EM}
{\mathcal L} = \frac{1}{2 \k^2} \left(R + \frac{6}{L^2} \right) - \frac{1}{4 e^2} F_{\mu\nu} F^{\mu\nu} \,.
\ee
Here $\k$ and $e$ are respectively the Newtonian and Maxwell constants while $L$ sets the cosmological constant lengthscale.

There is a unique regular solution to the theory (\ref{eq:EM}) with electric flux at infinity and that has rotations and spacetime translations as symmetries. This is the planar Reissner-Nordstr\"om-AdS black hole, with metric
\be
ds^2 = \frac{L^2}{r^2} \left(- f(r) dt^2 + \frac{dr^2}{f(r)} + dx^2 + dy^2 \right) \,.
\ee
The metric function here is
\be
f(r) = 1 - \left(1 + \frac{r_+^2 \mu^2}{2 \g^2} \right) \left(\frac{r}{r_+}\right)^{3} + \frac{r_+^2 \mu^2}{2 \g^2} \left(\frac{r}{r_+}\right)^{4} \,.
\ee
We introduced the dimensionless ratio of the Newtonian and Maxwell couplings
\be
\g^2 = \frac{e^2 L^2}{\k^2} \,.
\ee
The Maxwell potential of the solution is
\be
A = \mu \left(1 - \frac{r}{r_+} \right) dt \,.
\ee
We have required the Maxwell potential to vanish on the horizon, $A_t(r_+)=0$. The simplest argument for this condition is that otherwise the holonomy of the potential around the Euclidean time circle would remain nonzero when the circle collapsed at the horizon, indicating a singular gauge connection. The planar Reissner-Nordstr\"om-AdS solution is characterized by two scales, the chemical potential $\mu = \lim_{r \to 0} A_t$ and the horizon radius $r_+$.  From the dual field theory perspective, it is more physical to think in terms of the temperature than the horizon radius
\be
T = \frac{1}{4 \pi r_+} \left(3 - \frac{r_+^2 \mu^2}{2 \g^2} \right) \,.
\ee
The black hole is illustrated in figure \ref{fig:RN} below.
\begin{figure}[h!]
\begin{center}
\includegraphics[scale=.3]{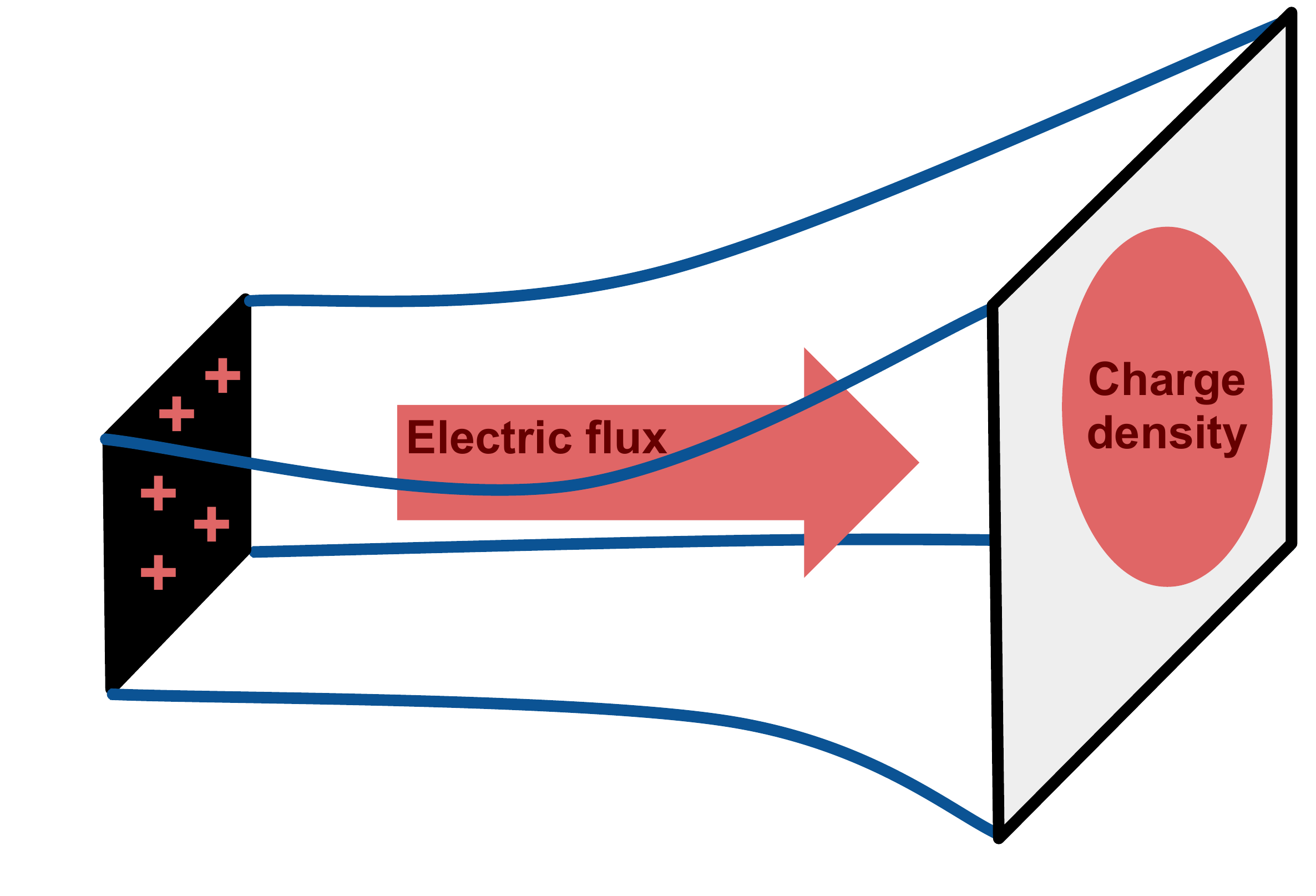}
\end{center}
\vspace{-.5cm}
\caption{
The planar Reissner-Nordstr\"om-AdS black hole. The charge density is sourced entirely by flux emanating from the black hole horizon.\label{fig:RN}}
\end{figure}
This black hole, which can additionally carry a magnetic charge, was the starting point for holographic approaches to finite density condensed matter \cite{Hartnoll:2007ih,Hartnoll:2007ip}.

Because the underlying UV theory is scale invariant, the only dimensionless quantity that we can discuss is the ratio $T/\mu$. In order to answer our basic question about the IR physics at low temperature, we must take the limit $T/\mu \ll 1$ of the solution. We thereby obtain the extremal Reissner-Nordstr\"om-AdS black hole with
\be
f(r) = 1 - 4 \left(\frac{r}{r_+}\right)^{3} +3 \left(\frac{r}{r_+}\right)^{4} \,.
\ee
The near-horizon extremal geometry, capturing the field theory IR, follows by expanding the solution near $r = r_+$. Setting $r = r_ +(1 - r_+/\rho)$, taking $\rho$ large and rescaling $\{t,x,y\}$ by dimensionless constants gives
\be\label{eq:RB}
ds^2 = \frac{L^2}{6} \left(\frac{- d\bar t^2 + d\rho^2}{\rho^2} \right) + d\bar x^2 + d\bar y^2 \,, \qquad A = \frac{\g}{\sqrt{6}} \frac{d\bar t}{\rho} \,.
\ee
The near horizon geometry is seen to be the famous Bertotti-Robinson-like spacetime $AdS_2 \times {\mathbb R}^2$ \cite{Gibbons}.

Consistently with the field theory intuition following from our discussion in section \ref{sec:motivation} above, the interior geometry (\ref{eq:RB}) exhibits an emergent IR scaling invariance
\be\label{eq:inftyscaling}
\rho \to \lambda \rho \,, \quad t \to \lambda t \,, \quad \{x,y\} \to \, \{ x,y\} \,.
\ee
Under this scaling, however, time scales but space does not. We will see shortly that such scaling is a degenerate limit of possible scalings that are anisotropic in time and space, with dynamical critical exponent $z=\infty$. This fact is directly related to the following two statements. Firstly: The entropy density remains finite at zero temperature
\be\label{eq:s0}
s = \frac{S}{V_2} = \frac{2 \pi}{\k^2} \frac{A}{V_2} = \frac{\pi \mu^2}{3 e^2} \,. \qquad (T=0).
\ee
Here $V_2$ is the field theory spatial volume, $A$ is the event horizon area, and the second equality is Hawking's formula for black hole entropy. Secondly: the density of states is IR divergent \cite{Jensen:2011su}. The IR scaling (\ref{eq:inftyscaling}) implies $ \rho(E) \sim e^{S} \, \d(E) + E^{-1}$. The first term is the zero temperature entropy density, while the second gives an IR divergence upon integrating.

The implication of the previous paragraph is that the near horizon geometry of extremal Reissner-Nordstr\"om-AdS black holes is unlikely to survive beyond the bulk classical `large $N$' limit. The entropy density suggests a fine tuning while the divergent number of states suggests an instability. In the following sections we will discuss two circumstances in which the near horizon $AdS_2 \times {\mathbb R}^2$ has instabilities at leading order in large $N$. The first involves bosons and leads to superconductivity, the second involves fermions. We then discuss the effects of incorporating dilaton couplings; in these cases $AdS_2 \times {\mathbb R}^2$ is not even a solution to the equations of motion. The new IR geometry in all cases captures physics analogous to Landau damping with finite $z$.

\section{Holographic superconductors}

Despite the concerns at the end of the previous section, a near horizon $AdS_2$ geometry is a robust feature of extremal black holes, see e.g. \cite{Sen:2005wa}. One way to evade this conclusion
is to add charged matter fields to our Einstein-Maxwell theory. This is a natural extension to consider; a consistent microscopic theory in the bulk will certainly have fields carrying the Maxwell charge. We will find that extremal black holes can become thermodynamically disfavored relative to new solutions to the equations of motion in which the asymptotic electric flux is explicitly supported by charged fields rather than emanating from behind an extremal horizon. If the charged matter is bosonic we obtain so-called `holographic superconductors' \cite{Hartnoll:2008vx, Hartnoll:2008kx} while in the case of charged fermions we obtain `electron stars' \cite{Hartnoll:2009ns, Hartnoll:2010gu}. Consider first the bosonic case. A simple Lagrangian density to consider is
\be\label{eq:EMH}
{\mathcal L} = \frac{1}{2 \k^2} \left(R + \frac{6}{L^2} \right) - \frac{1}{4 e^2} F_{\mu\nu} F^{\mu\nu} - |\nabla \phi - i A \phi|^2 - m^2 |\phi|^2 - V(|\phi|) \,.
\ee
We have taken the charge of the scalar to be unity, without loss of generality.

The question is whether there are circumstances under which it is favorable for the charged scalar field $\phi$ to condense. The Maxwell potential $A_t$ of the Reissner-Nordstr\"om background acts like a space-dependent chemical potential for the scalar. We might therefore anticipate Bose-Einstein condensation of $\phi$ \cite{Gubser:2008px}. Counteracting this possibility, we can recall that the gravitational well of asymptotically $AdS$ spacetimes acts like a covariant box and is capable of stabilizing potential tachyons \cite{Breitenlohner:1982bm}. Furthermore, the propensity of matter to fall into black hole horizons can be formalised into `no hair' theorems forbidding the presence of scalar fields outside of black holes in a range of circumstances, e.g. \cite{Hertog:2006rr}.

The conditions for instability simplify at zero temperature. Here stability is determined by the behaviour of the scalar field in the near horizon $AdS_2 \times {\mathbb R}^2$ geometry \cite{Hartnoll:2008kx, Gubser:2008pf, Denef:2009tp}. The scalar field will condense if its effective mass squared in the near horizon region is below the $AdS_2$ Breitenlohner-Freedman bound ($m^2_{\text{B.F}} = -1/(4 L_{AdS_2}^2)$, see e.g. \cite{Klebanov:1999tb})
\be\label{eq:ineq}
L_{AdS_2}^2 (m^2 + g^{tt} A_t A_t) = \frac{1}{6} \left(m^2 L^2 -\g^2 \right) \leq -\frac{1}{4} \,.
\ee
Here we used the near horizon solution (\ref{eq:RB}). This expression can be understood physically as the condition for the classical field analogue of Schwinger pair production in $AdS_2$ \cite{Pioline:2005pf}. The spacetime becomes unstable to production of pairs of classical waves, as illustrated in the following figure \ref{fig:instability}.
\begin{figure}[h!]
\begin{center}
\includegraphics[scale=.3]{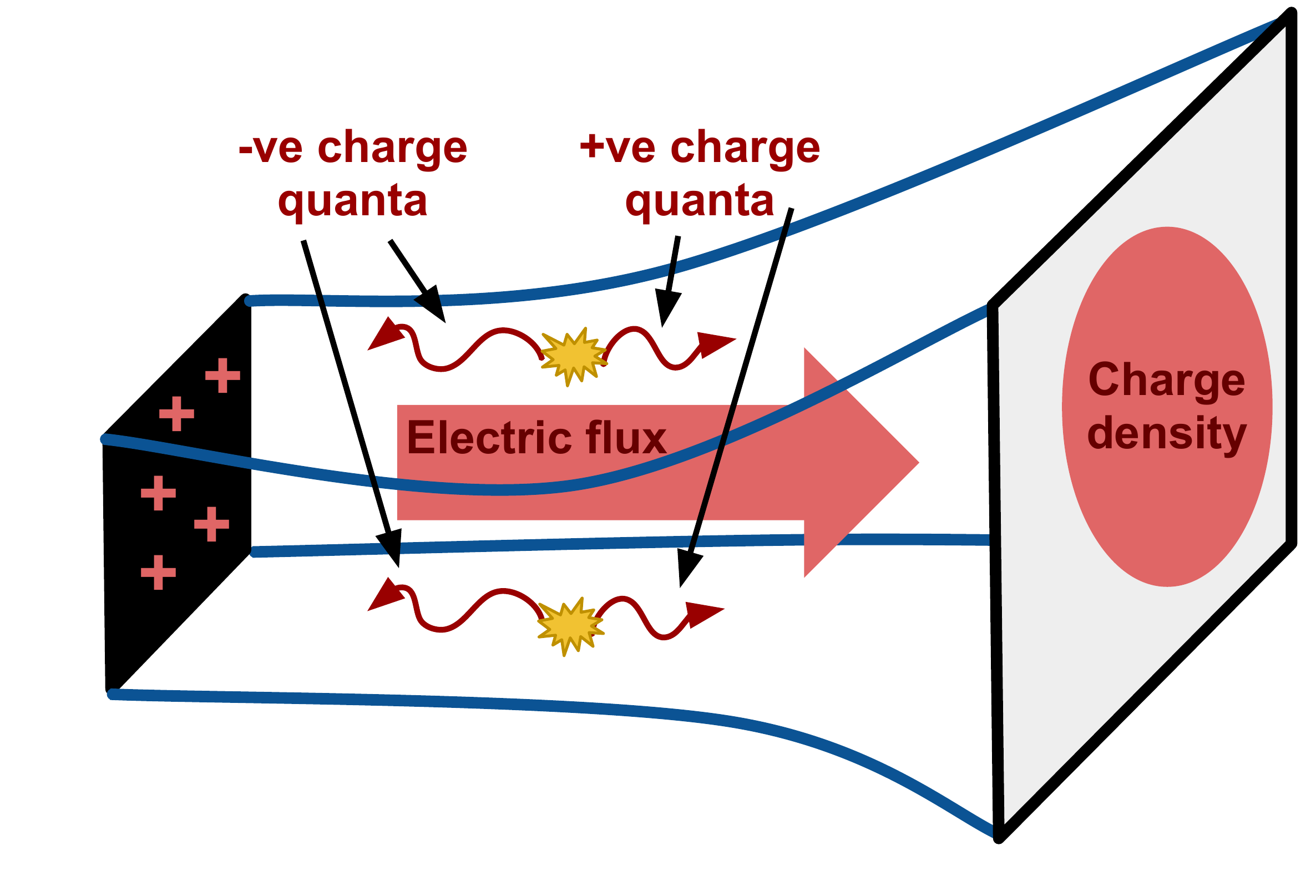}
\end{center}
\vspace{-.5cm}
\caption{
The onset of the superconducting instability. Pair produced waves are repelled or attracted to the black hole depending on their charge.\label{fig:instability}}
\end{figure}
The physics here can also be understood as the charged analogue of superradiance. It is more than an analogy, as Reissner-Nordstr\"om-AdS black holes can be uplifted to higher dimensional rotating black holes \cite{Chamblin:1999tk}. Superradiance leads to a genuine instability in the present context as the $AdS_4$ `box' reflects the waves back towards the black hole, leading to exponential growth of the mode.

Rather than the dynamical evolution of the instability, at fixed energy, we are interested in determining the dominant saddle point of the asymptotically $AdS_4$ Euclidean path integral at fixed temperature and chemical potential. It has been shown that in the model (\ref{eq:EMH}), if the mass satisfies the inequality (\ref{eq:ineq}), then it is thermodynamically favourable for the scalar field $\phi$ to condense below a critical temperature $T < T_C \propto \mu$. The condensation occurs via a second order phase transition \cite{Hartnoll:2008vx, Hartnoll:2008kx}. The condensate describes a macroscopically occupied bosonic ground state with a definite phase. It thereby spontaneously breaks the $U(1)$ symmetry of the theory. For this reason these solutions are known as holographic superconductors.

The objective of this chapter is to characterize phases of matter with gravity duals at low temperatures $T/\mu \ll 1$. We therefore omit a review of the many properties of holographic superconductors to focus on their low energy behaviour at low temperatures. The interested reader is referred to the discussions in \cite{Hartnoll:2009sz,Herzog:2009xv,Horowitz:2010gk}.

The problem becomes to determine the IR behavior (i.e. in the far interior of the spacetime) of solutions to the equations of motion following from (\ref{eq:EMH}). This behavior depends strongly on the choice of potential $V$. The cleanest set of behaviors to consider are those in which the scalar field $\phi$ tends to a constant $\phi_\infty$ in the IR. Preserving rotational and spacetime translational invariance implies that the holographic superconductor spacetime, Maxwell potential and scalar may be taken to have the general form
\be\label{eq:spacetime}
\frac{1}{L^2} ds^2 = - f(r)dt^2 + g(r) dr^2 + \frac{dx^2 + dy^2}{r^2} \,, \quad A = \g \, h(r) dt \,, \quad \phi = \phi(r) \,.
\ee
In these coordinates, a zero temperature IR spacetime without a finite size horizon will be at $r \to \infty$. If the scalar field does not stabilise as $r \to \infty$, as occurs for instance for $m^2 < 0$ and no potential $V$ \cite{Horowitz:2009ij}, then corrections to the leading order gravity action (\ref{eq:EMH}) will be necessary in order to access the true far IR of the theory at zero temperature. When the scalar does stabilise we can hope for a scale invariant solution. Suppose we find such an IR fixed point. A useful quantity to consider is the IR scaling dimension of the dual field theory charge density operator $J^t$, dual to $A_t$ in the bulk \cite{Gubser:2009cg}. The renormalisation group flow described by the spacetime (\ref{eq:spacetime}) is being driven by the UV insertion of $\mu J^t$. Lorentz invariance is broken along this flow and the bulk $U(1)$ Maxwell symmetry is Higgsed. The operator $J^t$ can therefore typically acquire an anomalous dimension. If it becomes irrelevant in the IR theory then we can expect that Lorentz symmetry will be restored and we will obtain an emergent $AdS_4$ spacetime. This phenomenon has been observed in various models, starting with \cite{Gubser:2008wz} and including cases that can be uplifted to consistent nonlinear solutions of string theory \cite{Gauntlett:2009dn, Gubser:2009gp}.

\begin{figure}[h!]
\begin{center}
\includegraphics[scale=.3]{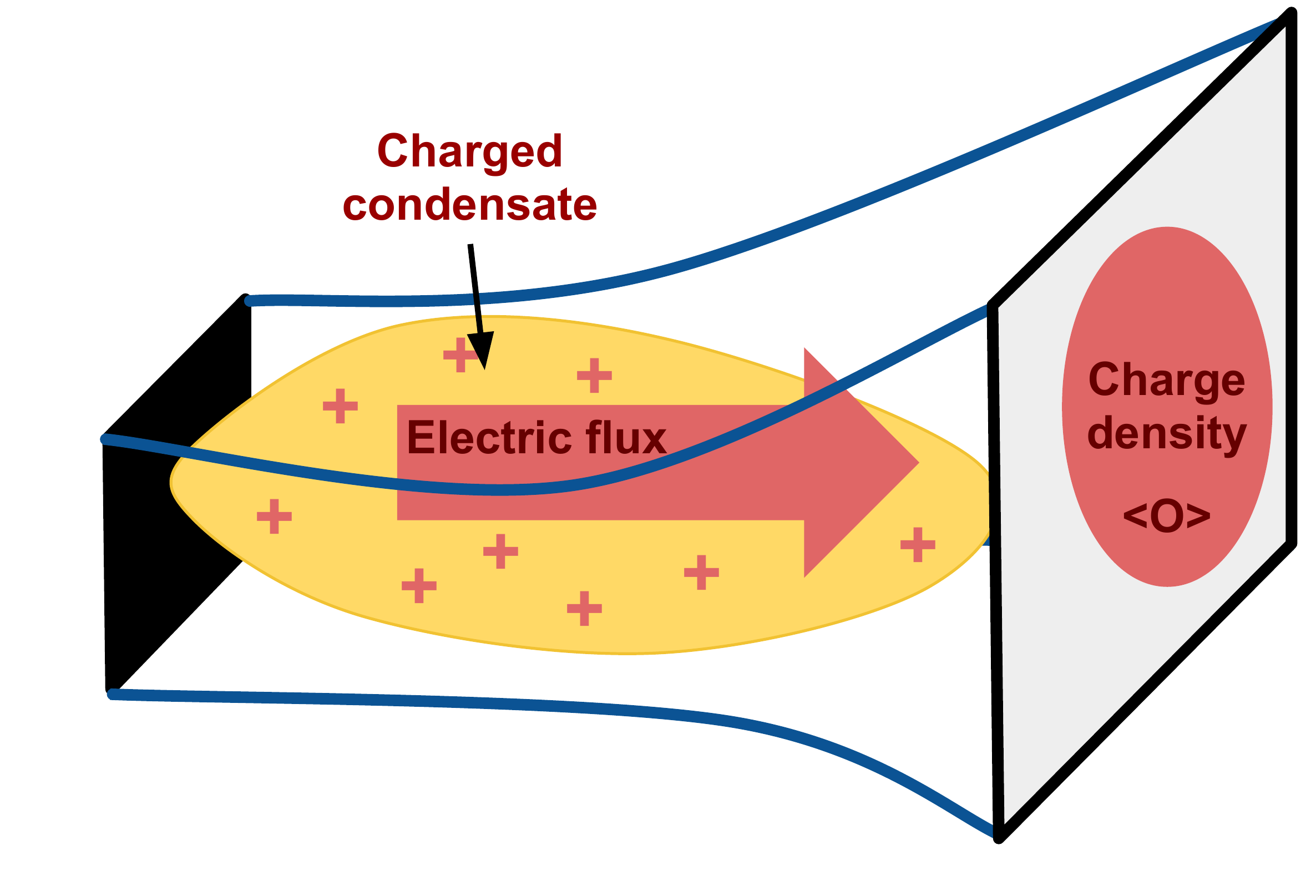}
\end{center}
\vspace{-.5cm}
\caption{
The zero temperature holographic superconductor. The electric flux is sourced entirely by the scalar field condensate.\label{fig:endpoint}}
\end{figure}

More generally, the operator $J^t$ will not be irrelevant in the IR and we have reason to anticipate a non-Lorentz invariant fixed point. Correspondingly, in \cite{Gubser:2009cg, Horowitz:2009ij} so-called `Lifshitz' \cite{Kachru:2008yh} geometries were found in the far IR. These take the form
\be\label{eq:lif}
\frac{1}{L^2}  ds^2 = - \frac{dt^2}{r^{2z}}  + g_\infty \frac{dr^2}{r^2} + \frac{dx^2 + dy^2}{r^2} \,, \quad A = \g \, h_\infty \frac{dt}{r^z} \,, \quad \phi = \phi_\infty \,.
\ee
Without loss of generality we have rescaled the time coordinate to remove any constant term in $g_{tt}$.
The Lifshitz solution has the scaling symmetry
\be\label{eq:zscaling}
r \to \lambda r \,, \quad t \to \lambda^z t \,, \quad \{x,y\} \to \lambda \{x,y\} \,.
\ee
The variable $z$ is called the dynamical critical exponent. Setting $r = \rho^{1/z}$ in the Lifshitz geometry (\ref{eq:lif}) and then comparing with the $AdS_2 \times {\mathbb R}^2$ near horizon geometry of the extremal black hole (\ref{eq:RB}), we see that the extremal black hole corresponds to the limit $z \to \infty$. The relativistic case of $AdS_4$ is $z=1$. An immediate property of the solution (\ref{eq:lif}) is that, unlike for the $AdS_2 \times {\mathbb R}^2$ case, there is no electric flux emanating from the horizon, ${\displaystyle \lim_{r \to \infty}} \int_{{\mathbb R}^2} \star F = 0$. All of the field theory charge density is therefore sourced by the condensate in the bulk, as we illustrate in figure \ref{fig:endpoint}.

Substituting the Lifshitz ansatz (\ref{eq:lif}) into the equations of motion, one finds that the theory (\ref{eq:EMH}) admits Lifshitz solutions with the dynamical critical exponent $z$ given by solutions to
\be\label{eq:zsol}
8(V_T - 3) + 4(V_T^{' \, 2} - 4 V_T + 12) z + (V_T^{' \, 2} + 8 V_T - 24) z^2 + V_T^{' \, 2} z^3 = 0 \,.
\ee
Here we introduced
\be
V_T = \k^2 L^2 \left(V(\phi_\infty) + m^2 \phi_\infty^2 \right) \,, \qquad V_T^{' } = \frac{\k^2 L}{e} \left(V'(\phi_\infty) + 2 m^2 \phi_\infty \right) \,.
\ee
Thus the dynamical critical exponent is determined by the value of the potential and its first derivative at the fixed point value of $\phi_\infty$, which is in turn determined by the equations of motion. In order for the scaling (\ref{eq:zscaling}) to have a straightforward interpretation as a renormalisation transformation, one should have $z>0$. The null energy condition in the bulk furthermore implies $z > 1$ \cite{Gubser:2009cg}. Even if (\ref{eq:zsol}) gives physical solutions for $z$, it is not guaranteed that the corresponding Lifshitz solution is realised as the near horizon geometry. An instructive simple case to consider is $m^2 > 0$ and $V=0$. One obtains in this case \cite{Gubser:2009cg, Horowitz:2009ij}
\be\label{eq:justm}
z = \frac{\g^2}{\g^2 - L^2 m^2} \,, \qquad \phi_\infty^2 = \frac{1}{e^2 L^2} \frac{6 z}{(1+z) (2+z)} \,.
\ee
The Lifshitz solutions are seen to exist so long as the scalar is not too heavy, $L^2 m^2 < \g^2$.  As $L^2 m^2 \to 0$, we see that $z \to 1$ and an emergent relativistic $AdS_4$ is obtained. As 
$L^2 m^2 \to \g^2$ from below, $z \to \infty$ and the extremal $AdS_2 \times {\mathbb R}^2$ geometry is recovered. However, recall from (\ref{eq:ineq}) that $AdS_2 \times {\mathbb R}^2$ is stable against $\phi$ condensing if $\g^2 - m^2 L^2 \leq \frac{3}{2}$. Extremal Reissner-Nordstr\"om is likely the ground state in this case. It follows that the Lifshitz geometries (\ref{eq:justm}) realized as IR scaling regimes in this theory with a positive quadratic potential have at most
\be
1 \leq z \leq \frac{2}{3} \g^2 \,.
\ee
Thus the dynamical critical exponent is bounded by the relative strength of the Maxwell and Newtonian couplings.

A few words about the near horizon Lifshitz geometry (\ref{eq:lif}) are appropriate at this point. As befits a scale invariant solution, all local curvature invariants constructed from the Riemann tensor are constant and small in Planck units when $\kappa/L \ll 1$ \cite{Kachru:2008yh}. Furthermore, the geometry is robust against higher derivative corrections, which can only renormalize the overall curvature scale $L$ and the dynamical critical exponent $z$ \cite{Adams:2008zk}. Nonetheless, Lifshitz spacetimes are geodesically incomplete. For $z \neq 1$, an infalling observer experiences divergent tidal forces as $r \to \infty$ \cite{Hartnoll:2009sz, Copsey:2010ya}.
Thus the spacetime strictly does not end at a horizon in the interior, but rather a null singularity. In the classical gravity limit $\kappa/L \ll 1$, a parametrically small nonzero temperature $0 < T/\mu \ll 1$ will lead to a regular finite temperature horizon, and is sufficient to safely bound all observables. In practice this will suffice for many purposes. However, the spirit of the investigations in this chapter is to access the lowest possible energy scales in the dual field theory. In this sense we would like to be able to set $T=0$ and probe all the way into the interior.
It remains to be seen, then, whether this singularity has interesting consequences for the IR of the dual field theory, via for instance the overproduction of excited string states near the singularity due to the large tidal forces \cite{Horowitz:1989bv}.

The emergent low energy Lifshitz scaling (\ref{eq:zscaling}) must of course remind us of the field theory scaling (\ref{eq:z3}). The holographic superconductor setting is different to that field theory case because it describes a symmetry breaking phase. In the holographic superconductor it is the bosonic superfluid density, rather than a density of fermions, that is `Landau damping' the transverse gauge fields of the dual field theory and leading to the Lifshitz scaling (\ref{eq:zscaling}). The Lifshitz IR geometry indicates that as well as the necessary Goldstone boson, due to spontaneous symmetry breaking, there are additional (neutral) gapless excitations at the lowest energy scales. We are arguing that these are (a gauge invariant version of) Landau damped $SU(N)$ gauge fields. This presence of gapless degrees of freedom is seen for instance in the power law scaling of the entropy density at low temperatures. Dimensional analysis and the Lifshitz scaling (\ref{eq:zscaling}) imply that the entropy density scales with temperature as follows\be\label{eq:sscaling}
s \propto T^{2/z} \,.
\ee
The coefficient of proportionality is large in the bulk classical limit.
This scaling can be verified by explicit construction of low temperature black hole solutions and is consistent with the previous observation in (\ref{eq:s0}) that $z = \infty$ leads to a finite zero temperature entropy density.
The Goldstone boson itself does not appear in the geometry in the leading classical limit, as it is a single mode compared with the large $N$ number of gauge fields. At one loop level in the bulk, fluctuations of the Goldstone mode leads to IR divergences in our case of a 2+1 dimensional field theory. These randomize the phase of the condensate and restore the $U(1)$ symmetry in accordance with the Coleman-Mermin-Wagner theorem \cite{Anninos:2010sq}. A general framework for incorporating the low energy physics of the Goldstone mode may be found in \cite{son}.

\section{Electron stars}

Charged fermions in the bulk can also qualitatively alter the interior of zero temperature, finite charge density spacetimes. The physics is similar to that of the charged bosons we have just discussed, but also has important differences. Pauli exclusion implies that fermions cannot macroscopically occupy their ground state. The state will therefore not have a coherent phase and the $U(1)$ symmetry remains unbroken. As is familiar from solid state physics, the presence of a sufficiently large background Maxwell potential $A_t$ does not cause Bose-Einstein condensation, but rather the buildup of a Fermi surface. The filled Fermi sea is a specific quantum vacuum of the fermions in the presence of a chemical potential, stable in absence of interactions. Gravitating Fermi surfaces are also familiar, they are the neutron stars of astrophysics. The solutions we discuss in the following are charged, planar cousins of neutron stars, and therefore we call them electron stars.

Consider a free, charged Dirac field added to the Einstein-Maxwell action
\be\label{eq:EMD}
{\mathcal L} = \frac{1}{2 \k^2} \left(R + \frac{6}{L^2} \right) - \frac{1}{4 e^2} F_{\mu\nu} F^{\mu\nu} - \bar \psi \Gamma \cdot \left(\pa + \qtr \w_{\mu\nu}\G^{\mu\nu} - i A \right) \psi - m^2 \bar \psi \psi \,.
\ee
Here $\G^{\mu\nu}$ is an antisymmetrised gamma matrix and $\w_{\mu\nu}$ the spin connection. As we did for charged scalars in the previous section, we can ask after the fate of extremal Reissner-Nordstr\"om black holes in the presence of the charged fermion. Because fermion statistics prevents macroscopic occupation of states, fermionic instabilities should not be seen as an exponentially growing classical solution to the Dirac equation. Indeed such classical instabilities do not occur \cite{Faulkner:2009wj}. However, Schwinger pair production of fermions will occur in the near horizon $AdS_2 \times {\mathbb R}^2$ geometry if the fermion mass is sufficiently low \cite{Pioline:2005pf}
\be\label{eq:ineqF}
m^2 L^2 \leq \g^2 \,.
\ee
This is the fermionic analogue of the inequality (\ref{eq:ineq}) for bosons. In the fermionic case, while satisfying this inequality does not lead to a classical instability, it is manifested in solutions to the Dirac equation by the fermion acquiring an imaginary scaling dimension in the $AdS_2 \times {\mathbb R}^2$ spacetime \cite{Liu:2009dm, Faulkner:2009wj}. Analogously to the discussion for bosons, illustrated in figure \ref{fig:instability}, we might expect that pair production will lead to neutralisation of the black hole and a Fermi sea outside the horizon carrying the charge.

A further, logically separate, indication that a Fermi sea will become populated and backreact on the spacetime is the presence of Fermi surface singularities in fermion Green's functions in the extremal Reissner-Nordstr\"om background \cite{Lee:2008xf, Liu:2009dm, Cubrovic:2009ye, Faulkner:2009wj}. In the textbook case of fermions in flat space with a constant chemical potential, the residue of the Fermi surface singularity in the Green's function is related to a density of fermions via the Migdal relation. A similar relation should be anticipated in our curved spacetime \cite{Cubrovic:2010bf}, where now the density of fermions will gravitate.

Extremal Reissner-Nordstr\"om-AdS can be quantum mechanically unstable to the above two mechanisms populating the Fermi sea. This is the statement that the unpopulated extremal Reissner-Nordstr\"om fermion vacuum is thermodynamically unstable. We will show this by constructing solutions with a populated Fermi sea and finding that they have a lower free energy.

It is in general difficult to find solutions to the Einstein-Maxwell-Dirac system (\ref{eq:EMD}) with a populated and gravitating Fermi sea. This is because one must find all the eigenstates of the Dirac operator in a given background with energy below the chemical potential, sum their contributions to the bulk energy-momentum tensor, and then self-consistently backreact this energy-momentum tensor on the geometry to solve the equations of motion. The problem simplifies in a coarse-grained limit in which the fermions may be treated as an ideal fluid \cite{Hartnoll:2009ns, Hartnoll:2010gu}. In our charged and gravitating context, this can be called the Thomas-Fermi-Oppenheimer-Volkov limit. Mathematically, the limit is a WKB limit in which the Dirac eigenstates become very localised in the geometry and are therefore locally not sensitive to variations of the curvature and Maxwell field.

The WKB electron star limit requires the mass $m$ of the fermion to be large in units of the overall curvature lengthscale $L$ of the spacetime. Furthermore, a consistent solution requires the attractive gravitational and repulsive electrostatic forces between fermions making up the star to be comparable. Thus we impose
\be\label{eq:wkb}
m L \sim \frac{e L}{\k} \equiv \g \gg 1 \,.
\ee
In addition to this WKB limit, by working in the classical limit throughout this chapter we have already been assuming that the Maxwell and Newton couplings are small: $e \ll 1$ and $\k/L \ll 1$. 
The notion that a large density of noninteracting fermions with Compton wavelength much smaller than the scale at which the background varies can be described as an ideal fluid is intuitively plausible and so we will not support this claim here. The process of coarse graining is discussed from various angles in \cite{Ruffini:1969qy, deBoer:2009wk, Hartnoll:2010gu, Arsiwalla:2010bt, Hartnoll:2011dm}.

The action for a charged, gravitating ideal fluid can be written
\be\label{eq:actp}
{\mathcal L} = \frac{1}{2 \k^2} \left(R + \frac{6}{L^2} \right) - \frac{1}{4 e^2} F_{\mu\nu} F^{\mu\nu} + p(\mu_\text{loc.},s) \,.
\ee
This is Schutz's form of the action for an ideal fluid \cite{Schutz:1970my}, generalised to allow the fluid to be charged \cite{Hartnoll:2010gu}. To obtain the expected ideal fluid equations of motion, one must write the local chemical potential 
$\mu_\text{loc.} =  |d\phi + \alpha d\beta + \theta ds + A |$, and then vary with respect to the fluid potential variables $\{\phi, \alpha, \beta\}$ as well as the local entropy density $s$ and `thermasy' $\theta$. We will only consider zero temperature irrotational fluids where $\a = \b = \theta = s = 0$ and $\phi$ can be absorbed into $A$ by a gauge transformation. The potential $A$ will only have a time component. Thus we have that
\be\label{eq:muloc}
\mu_\text{loc.} = \frac{A_t}{\sqrt{g_{tt}}} \,.
\ee
This is of course simply the chemical potential as seen in the local rest frame of the fermions.
The pressure $p$ is determined by the local chemical potential via the charge density $\sigma$ and the energy density $\rho$
\be\label{eq:p}
- p = \rho - \mu_\text{loc.} \sigma \,, \qquad \rho = \int_m^{\mu_\text{loc.}} E g(E) dE \,,
\qquad  \sigma = \int_m^{\mu_\text{loc.}} g(E) dE \,,
\ee
where the density of states
\be
g(E) = \frac{1}{\pi^2} E \sqrt{E^2 - m^2} \,.
\ee

To find the electron star solution, one makes the same ansatz (\ref{eq:spacetime}) as above for holographic superconductors. Now there is no scalar field $\phi$, but rather the fluid tracks the background metric and Maxwell fields via (\ref{eq:p}). It is straightforward  to solve the Einstein-Maxwell-ideal fluid equations of motion numerically to find the profile of the star \cite{Hartnoll:2010gu}. Before concentrating on the interior IR geometry, we can make two qualitative comments. Firstly, it is clear from the equations of state (\ref{eq:p}) that the fluid is present only if the local chemical potential (\ref{eq:muloc}) is larger than the rest mass energy of the fermions
\be\label{eq:cond}
\mu_\text{loc.} > m \,.
\ee
This is the condition to populate the local Fermi sea. Looking for extremal solutions, without a finite temperature horizon, one finds that the condition (\ref{eq:cond}) is satisfied from the deep IR up to a specific radius. This radius is the boundary of the star. Outside of the star, the geometry becomes Reissner-Nordstr\"om-AdS with a mass and charge determined by integrating over the fermions in the star. This type of solution is illustrated in figure \ref{star} below.
\begin{figure}[h!]
\begin{center}
\includegraphics[scale=.3]{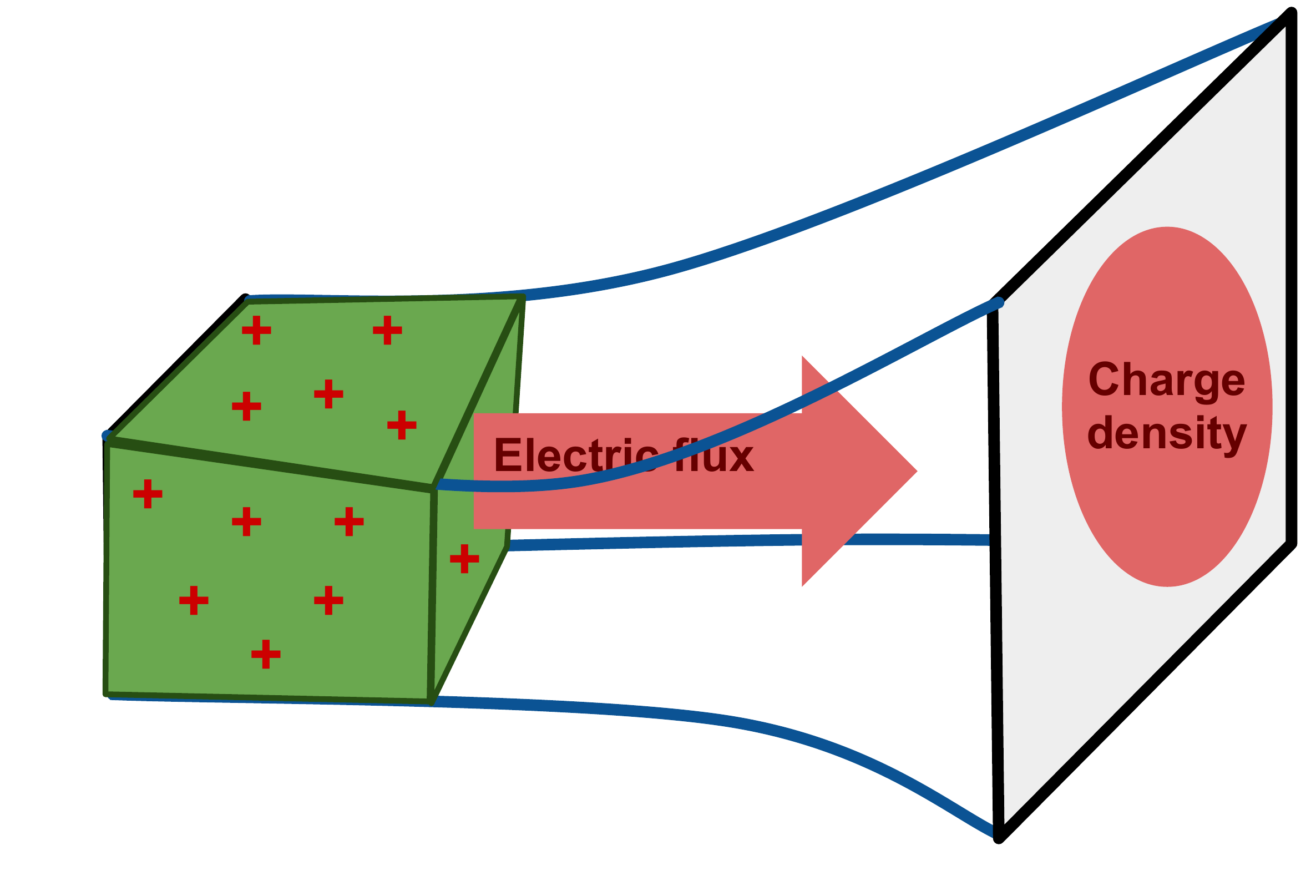}
\end{center}
\vspace{-.5cm}
\caption{
The electron star. The electric flux is sourced entirely by a fluid of fermions. The fluid is present at all radii for which the local chemical potential is greater than the fermion mass.\label{star}}
\end{figure}
Analogously to the holographic superconductors, at zero temperature all the charge is carried by the fermions rather than lying behind a horizon.

Secondly, consider heating up the system. At leading order in the semiclassical limit, this means placing a finite temperature black hole horizon in the interior of the spacetime. The fluid remains at zero temperature, as the fluid must be in thermal equilibrium with the Hawking radiation of the black hole, the effects of which are negligible in the semiclassical limit. At finite temperature, a fraction of the charge is carried by the black hole horizon, which subsequently pushes the fermion fluid a finite distance away from the horizon. The star becomes a band of fluid with an inner and an outer radius \cite{Hartnoll:2010ik, Puletti:2010de}. 
At nonzero temperature we can dial the ratio $T/\mu$. We can expect that at sufficiently high values of this ratio, the star will collapse to form a black hole. This will be analogous to the maximal mass of spherical neutron stars; in global rather than planar $AdS$, the mass scale can be compared to the radius of the spatial boundary sphere. In that case there is a first order phase transition: above a critical mass the degeneracy pressure cannot sustain the star \cite{deBoer:2009wk, Arsiwalla:2010bt}. In our planar setup we might anticipate a second order transition similar to that of the holographic superconductors above. In fact, the transition turns out to be third order \cite{Hartnoll:2010ik, Puletti:2010de}. The reason for the softness of the transition is that the free energy of the fluid is given by its pressure, and the pressure turns on relatively slowly when $\mu_\text{loc.} = m + \d\mu$ is only slightly above the fermion mass: from the formulae in (\ref{eq:p}) one obtains $p \sim (\d \mu)^{5/2}$ \cite{Hartnoll:2010ik}. In contrast the energy and charge densities go like $\rho,\sigma \sim (\d \mu)^{3/2}$.

Electron stars are found to exist whenever the fermion mass satisfies the condition (\ref{eq:ineqF}) for Schwinger pair production to occur in the would-be near horizon $AdS_2 \times {\mathbb R}^2$ region. In these cases, for all temperatures $T < T_C \propto \mu$ the electron star has a lower free energy than the corresponding Reissner-Nordstr\"om-AdS black hole  \cite{Hartnoll:2010gu, Hartnoll:2010ik, Puletti:2010de}. It is thermodynamically favorable to populate the Fermi sea.

We now return to our main focus of the emergent IR geometry at zero temperature. Despite preserving the $U(1)$ symmetry, electron stars obey similar equations to holographic superconductors. We can again make a Lifshitz scaling ansatz (\ref{eq:lif}) for the $r \to \infty$ IR geometry. Substituting the ansatz (\ref{eq:lif}) into the equations of motion following from the action
(\ref{eq:actp}) gives the dynamical critical exponent $z$ in terms of the pressure and energy density in the interior of the star
\be\label{eq:zstar}
\frac{2(6+3 \hat p_\infty + \hat \rho_\infty)}{\hat p_\infty + \hat \rho_\infty} - \frac{12 + \hat p_\infty - 3 \hat \rho_\infty}{\hat p_\infty + \hat \rho_\infty} z + z^2 = 0 \,.
\ee
Here $\{\hat p, \hat \rho \} = L^2 \k^2 \{p, \rho \}$. This expression is general, it will continue to hold for a different equation of state for the fermion fluid than that of free fermions. The local chemical potential (\ref{eq:muloc}) in the interior of the star is found to be, again independently of the equation of state,
\be\label{eq:muint}
L^2 \mu_\infty^2 = \g^2 \frac{z-1}{z} \,.
\ee
Here we see that the star will have $z \geq 1$. For the case of free fermions, we can perform the integrals in (\ref{eq:p}) with the value (\ref{eq:muint}) for the local chemical potential. Substituting the results of the integrals into (\ref{eq:zstar}) gives a formula that can be solved numerically to obtain $z$. By scaling the integrals in (\ref{eq:zstar}) and using (\ref{eq:muint}) one finds that $\hat p, \hat \rho \sim e^2 \g^2$. Thus from (\ref{eq:zstar}), $z$ of order one requires $e^2 \g^2 \sim 1$. Curiously, this is a regime in which the gravitational coupling is the square of the Maxwell coupling, $e^4 \sim \k^2/L^2$, reminiscent of the relation between closed and open string couplings. The explicit dependence of $z$ on $e^2 \g^2$ can be found numerically \cite{Hartnoll:2010gu}. Somewhat analogously to the expression (\ref{eq:justm}) for a class of holographic superconductors, it is found that at fixed mass $m$, if $e^2 \g^2 \to \infty$ then $z \to 1$ from above, while if $e^2 \g^2 \to 0$ then $z \to \infty$. The possible interior geometries of electron stars range from $AdS_4$ to $AdS_2 \times {\mathbb{R}}^2$.

As previously with the holographic superconductors, it is gratifying to see the emergence of an IR scaling regime due to finite density at strong coupling that mirrors the one loop Landau damping physics discussed in section \ref{sec:motivation}. The fermions and bosons of the current and previous sections, respectively, can in fact be combined to give an emergent scaling partially due to a symmetry breaking condensate and partially to a density of fermions \cite{Edalati:2011yv}. One potential limitation of the perturbative computation was the controlled determination of $z$. In the gravitational framework we have explored, we see that $z$ is tied to the ratio of Newton and Maxwell couplings, and to the mass of the charged fields. In a field theoretic language, these quantities translate respectively into the `central charges' characterizing the two point function of the energy momentum tensor and electric current, and to the anomalous scaling dimension of the gauge invariant charged operator carrying the charge density.

\section{Dilatonic scalars: Lifshitz and beyond}
\label{sec:dilaton}
  
The bottom line following from the previous two sections is that whenever it is possible to pair produce fermions or bosons in the vicinity of a planar, extremal Reissner-Nordstr\"om-AdS horizon, then the extremal black hole is not the thermodynamically preferred zero temperature, finite charge density spacetime. The dominant ground states we have found are distinguished from extremal Reissner-Nordstr\"om-AdS by two properties. Firstly, all the electric flux is sourced explicitly by charged bosonic and fermionic fields rather than emanating from behind an event horizon.  Secondly, the emergent IR scaling geometry had a finite $z$ Lifshitz scaling, rather than the $z=\infty$ scaling of $AdS_2 \times {\mathbb R}^2$. We might ask whether it is possible to decouple these two effects, that is, whether it is possible to have an emergent finite $z$ scaling together with the flux being carried by a horizon rather than charged fields.

We can heuristically think of the $AdS_2 \times {\mathbb R}^2$ near horizon geometry as being pushed open by the flux it carries.
In the Lifshitz solutions of the previous sections, the flux is consumed by the charged fermions and bosons as we move towards the `horizon' and therefore the spatial metric is able to collapse. In the absence of charged sources, the solution has nowhere to dump the electric flux and so we need an alternate mechanism to violate Gauss's law. A simple way to do this is to introduce a dilaton field. A minimal Einstein-Maxwell-dilaton-AdS action takes the form
\be\label{eq:dil}
{\mathcal L} = \frac{1}{2 \k^2} \left(R + \frac{6}{L^2} \right) - \frac{1}{4 e^2} e^{2 \a \phi} F_{\mu\nu} F^{\mu\nu}  - \frac{1}{2} (\nabla \phi )^2 \,.
\ee
Without loss of generality, consider $\a > 0$. If the dilaton $\phi$ is not constant in the near horizon region, then the effective Maxwell coupling $e_\text{eff.} = e \, e^{- \a \phi}$ will continue running and we may be able to escape landing at an $AdS_2 \times {\mathbb R}^2$ attractor point.

Indeed it is found that the IR geometry of the dilaton theory (\ref{eq:dil}), with an asymptotic electric flux, takes the Lifshitz form (\ref{eq:lif}), with $z > 1$ generically \cite{Taylor:2008tg, Goldstein:2009cv}. The important difference relative to (\ref{eq:lif}) is that, as we anticipated, the dilaton is not constant but rather grows logarithmically as $r \to \infty$:
\be\label{eq:philog}
\phi = k \, \log r \,,
\ee
with $k > 0$. The effective Maxwell coupling therefore vanishes in the far interior. Within a string theoretic framework, this leads us to be concerned about large stringy effects in the near horizon geometry. Higher derivative corrections may be expected to stabilise the dilaton at a constant value, leading again to an $AdS_2 \times {\mathbb R}^2$ near horizon geometry \cite{Sen:2005wa}. If this occurs, the semiclassical gravitational limit has nonetheless provided a parametrically large window of IR energy scales controlled by a $z>1$ scaling. The dilaton itself is covariant rather than invariant under this scaling.

Once accustomed to the logarithmic running of the dilaton in (\ref{eq:philog}), one can expect that allowing couplings other than the Maxwell coupling to run with the dilaton may induce new effects in the near horizon geometry. For instance, consider a potential for the dilaton.
Assume that the potential is dominated by an exponential term in the near horizon region where the dilaton is growing according to (\ref{eq:philog}). Thus we write the Lagrangian density
\be\label{eq:dil2}
{\mathcal L} = \frac{1}{2 \k^2} \left(R - V_0 \, e^{2 \b \phi} \right) - \frac{1}{4 e^2} e^{2 \a \phi} F_{\mu\nu} F^{\mu\nu}  - \frac{1}{2} (\nabla \phi )^2  \,.
\ee
The cosmological constant term has been replaced by the potential $V_0 \, e^{2 \b \phi}$ that will no longer be constant in the near horizon limit. Note that the action (\ref{eq:dil2}) is only the action to leading order in a large dilaton. In order for the theory to have e.g. an asymptotic $AdS_4$ vacuum there must be additional terms in the potential to asymptotically stabilise the dilaton.

The near horizon scaling solutions of (\ref{eq:dil2}) take a more general form than the Lifshitz geometry. The metric must generically be written \cite{Gubser:2009qt, Charmousis:2010zz, Iizuka:2011hg}
\be\label{eq:beyondlif}
ds^2 = r^{2 \d} \left( - \frac{dt^2}{r^{2z}}  + g_\infty \frac{dr^2}{r^2} + \frac{dx^2 + dy^2}{r^2} \right) \,.
\ee
The extra overall term compared to (\ref{eq:lif}) implies that now the metric itself transforms covariantly under scalings rather than being invariant. This fact has immediate consequences for the thermodynamics of the system. For instance, while Lifshitz invariance implies the entropy density scaling of (\ref{eq:sscaling}) with temperature for all of the systems considered thus far, heating up the Lifshitz-covariant spacetime (\ref{eq:beyondlif}) leads to a low temperature dependence of the entropy density of \cite{Gubser:2009qt, Charmousis:2010zz, Iizuka:2011hg}
\be
s \propto T^{2 (1-\d)/z} \,.
\ee
This scaling can be recovered from a dimensional analysis that imparts an `anomalous' scaling dimensionality to the spatial volume, while keeping the relative scaling between space and time determined by $z$. Another consequence is that the curvature scalars of the spacetime (\ref{eq:beyondlif}) are no longer constant and one must worry once more about stringy and quantum gravity effects in the far IR of the geometry. The reader is referred to discussions in \cite{Charmousis:2010zz, Iizuka:2011hg}.

The summary of this section is that, within the semiclassical regime at least, dilaton spacetimes with an asymptotic electric flux have near horizon geometries characterized by a finite $z$ scaling symmetry. This is achieved without explicit charged matter. All the flux, as defined by $\int_{{\mathbb R}^2} \star \left(e^{2 \a \phi} F\right)$, emanates from behind the IR `horizon'.
 
\section{Horizons, fractionalization and black hole information}

In this last section we will tie together the various solutions discussed into an interpretational framework. To start, recall the correspondence between the basic ingredients at our disposal in the gravitational and field theoretic descriptions of the system. This is the connection between bulk fields and gauge invariant operators mentioned in section \ref{sec:holo} above. We can write
\be\label{eq:table}
\begin{array}{rcl}
\text{Gravity} & | & \text{Field theory} \\
g_{\mu\nu} & = & T^{\mu\nu} \\
A_\mu & = & J^\mu \\
\phi & \sim & \tr \left(\Phi \Phi\right), \, \tr \left(\Psi \Psi \right) \\
\psi & \sim & \tr \left(\Phi \Psi \right) \\
\phi \; \text{(dilaton)} & \sim & \tr F_{\mu\nu} F^{\mu \nu} \,.
\end{array}
\ee
For the last three rows in this table, the precise matching between gauge invariant observables and bulk fields depends on the particular instance of duality being considered. The schematic correspondence indicated for these cases shows roughly the types of simple operators that will correspond to our bulk fields. On the other hand, the `elementary' field theoretic operators $\Phi,\Psi, F_{\mu\nu}$ are not directly accessible in the bulk, as they are not gauge invariant. In a slight abuse of terminology, we will refer to the gauge invariant operators dual to $\phi,\psi$ as `mesonic'. The name is supposed to remind us that these operators are composite in terms of the fields appearing in a `microscopic' field theory Lagrangian. 

Next, some general comments about horizons. The physics of horizons in holography is overdetermined (in the Freudian rather than mathematical sense of the word); they seem to play multiple independent, crucial roles simultaneously. These range from providing a mechanism for dissipation at leading order in the large $N$ expansion to geometrically realizing, for instance, the IR temperature scale and associated physics such as thermal screening. Here we wish to argue that horizons at finite charge density have the additional role of enabling a gauge-invariant description of `fractionalized' phases where microscopically the electric charge would seem to be carried by gauge-charged operators. Very loosely speaking, the gauge-variance is hidden behind the horizon. The word `fractionalized' is used here in analogy with the condensed matter construction (\ref{eq:cfb}). We can think of the gauge invariant fermion $c$ as being analogous to the mesonic field $\psi$, while the gauge charged fermion and boson $f$ and $b$ are analogous to the `microscopic' fields $\Psi$ and $\Phi$.

The identification of horizons with deconfined phases, as defined for instance through an expectation value for the Polyakov loop, is a seminal result in holography \cite{Witten:1998zw}. The spirit of the physics here will be similar, except that we are primarily interested in the charged sector at zero temperature.

One route to understanding the ubiquity of horizons in holography is the following. Recall from section \ref{sec:prelim} above that two key ingredients of holography were, firstly, that the large $N$ limit enabled the correlators of certain `single trace' operators to factorize and, secondly, that there should only be a handful of such single trace operators with order one anomalous dimension. This second statement requires the theory to be strongly interacting as otherwise there will be an infinite tower of single trace operators with roughly evenly spaced dimensions. Taking these statements together one is led to the notion of a `generalised free field'. These are operators that factorize but which do not obey free wave equations in the field theory Minkowski space. Examples are the gauge-invariant operators appearing in (\ref{eq:table}). One can then show that generalized free fields can only arise as a small subsector of a larger theory \cite{ElShowk:2011ag}, consistent with the large central charge (\ref{eq:central}) in the holographic context. This means that to reconstruct the full theory, the set of gauge invariant operators we have been considering must be completed with a large number (order $N^2$, say) of operators with large anomalous dimensions. These operators will not (typically) have a classical geometric description in the bulk, but will rather appear as `black hole microstates'. The fact that this large number of `heavy' states, with fixed mass and charges, are indistinguishable to most bulk probes leads to the finite entropy density of black hole horizons. 

We see, therefore, that the elegant distinction between `mesonic' phases, where the flux is sourced by charged fields in the bulk dual, and `fractionalized' phases, where the flux is sourced by horizons, is made possible by the holographic large $N$ limit. This is because, as we just explained, this limit creates a hierarchy in the anomalous dimensions of operators in the theory, and it is this hierarchy that allows the distinction between black hole states and classical bulk fields. The same mechanism underlies holographic descriptions of
finite temperature deconfinement transitions.

Seemingly independently of holography, the mean field limit of the fractionalized Fermi liquid phase of the lattice Anderson model \cite{subir} was also found to lead to a zero temperature entropy density, and consequently an effective $AdS_2 \times {\mathbb R^2}$-like IR scaling regime. Taken together with the holographic results, this may suggest that, in `classical' limits, charge fractionalization is closely tied to a finite entropy density (a similar conclusion is argued for in \cite{Iqbal:2011in}). An obstacle to such a general conclusion are the charged dilatonic near horizon geometries of section \ref{sec:dilaton}. While the running dilaton of these solutions is presumably stabilised at an $AdS_2 \times {\mathbb R^2}$ attractor point once higher derivative corrections are included \cite{Sen:2005wa}, this is not visible in the leading classical bulk limit. The probable lesson here is that the generalised free field/black hole microstate dichotomy does not guarantee that there are sufficiently many charged black hole microstates to generate classical charged black hole horizons. Consequently the leading order in $N$ low energy physics need not have a $z=\infty$ scaling symmetry.

So far we have not defined fractionalization in a system-invariant way. One way to make the notion more precise is via the `Luttinger count' \cite{Huijse:2011hp}. Under quite general circumstances, compressible finite density phases of matter, in which the $U(1)$ symmetry is unbroken, are expected to obey the Luttinger theorem. This theorem says that the total charge density is equal to the sum over the momentum space volumes of all Fermi surfaces in the theory, weighted by the charge of the corresponding fermionic operators
\be\label{eq:luttinger}
\langle J^t \rangle = \sum_{\ell \in \text{fermions}} q_\ell V_\ell \,.
\ee
The Fermi surfaces are defined as the singular locii of the fermion Green's function at zero energy: e.g. $G^{-1}_\ell(k=k_F,\w=0)=0$.
An essential feature of a `fractionalized Fermi liquid'  \cite{fraction} is that while the Luttinger count (\ref{eq:luttinger}) remains true in the presence of gauge fields, the corresponding gauge-charged Fermi surfaces may not be detectable by gauge-invariant fermion probes of the system. Thus, when we sum over all Fermi surfaces of gauge-invariant operators, we may encounter a deficit in the Luttinger count. Thus the difference between the right and left hand sides of (\ref{eq:luttinger}), with the sum restricted to gauge-invariant mesonic fermions,
gives a measure of the fractionalized nature of the system.

Using the Luttinger count, let us compare a stable extremal Reissner-Nordstr\"om spacetime, i.e. with all charged bosons and fermions heavier than the bounds (\ref{eq:ineq}) and (\ref{eq:ineqF}), with an electron star. In a WKB limit for the fermions, one can establish that the field theory charge density equals the flux emanating from the horizon plus the charge carried by a fermion fluid outside the black hole, see e.g. \cite{Hartnoll:2010ik}. The bulk fermions obey a bulk Luttinger theorem at each radius \cite{Hartnoll:2010gu} and therefore one might anticipate that the flux emanating from the horizon will equal the `missing' contribution to the gauge-invariant Luttinger count. That is, we expect
\be
\begin{array}{c}
\text{Flux from} \\
\text{horizon} 
\end{array}
\; = \;  \langle J^t \rangle - 
\sum \begin{array}{c}
\text{Fermi surface} \\
\text{volumes of $\psi$} 
\end{array} \,.
\ee
This relation was made precise in \cite{Hartnoll:2011dm, Iqbal:2011in}, where it was shown that in field theory phases dual to an electron star, where all the charge is sourced by bulk fermions, the Luttinger equality (\ref{eq:luttinger}) holds when the sum is restricted to gauge-invariant Fermi surfaces. Conversely, an extremal Reissner-Nordstr\"om spacetime that is stable against WKB fermions condensing will have no associated gauge-invariant Fermi surfaces.\footnote{Away from the WKB limit there is a small window of bulk fermion charges and masses where it is possible to have a gauge-invariant Fermi surface without pair producing fermions near the horizon according to the criterion (\ref{eq:ineqF}). See figure 6 of \cite{Faulkner:2009wj}. In these cases, extremal Reissner-Nordstr\"om will coexist with a parametrically small amount of charge carried by fermions outside the horizon.}
The Luttinger relation (\ref{eq:luttinger}) is therefore maximally violated in this case. At the time of writing, only these limiting cases of the `fully mesonic' electrons stars and the `fully fractionalized' extremal black holes have been constructed. No doubt, before this book is published, intermediate cases will also have been realised holographically. The important point is that a deficit in the gauge-invariant Luttinger count gives a direct connection between charged horizons and fractionalized phases of matter.


\begin{thebibliography}{8}
  
\bibitem{Maldacena:1997re}
  J.~M.~Maldacena,
  ``The Large N limit of superconformal field theories and supergravity,''
  Adv.\ Theor.\ Math.\ Phys.\  {\bf 2}, 231-252 (1998).
  [hep-th/9711200].
  
\bibitem{Polchinski:2010hw}
  J.~Polchinski,
  ``Introduction to Gauge/Gravity Duality,''
    [arXiv:1010.6134 [hep-th]].
  
\bibitem{McGreevy:2009xe}
  J.~McGreevy,
  ``Holographic duality with a view toward many-body physics,''
  Adv.\ High Energy Phys.\  {\bf 2010}, 723105 (2010).
  [arXiv:0909.0518 [hep-th]]. 
    
\bibitem{Heemskerk:2010hk}
  I.~Heemskerk, J.~Polchinski,
  ``Holographic and Wilsonian Renormalization Groups,''
  [arXiv:1010.1264 [hep-th]].   

\bibitem{Faulkner:2010jy}
  T.~Faulkner, H.~Liu, M.~Rangamani,
  ``Integrating out geometry: Holographic Wilsonian RG and the membrane paradigm,''
  [arXiv:1010.4036 [hep-th]].
  
\bibitem{Heemskerk:2009pn}
  I.~Heemskerk, J.~Penedones, J.~Polchinski, J.~Sully,
  ``Holography from Conformal Field Theory,''
  JHEP {\bf 0910}, 079 (2009).
  [arXiv:0907.0151 [hep-th]]. 
  
  \bibitem{ed}
  E.~Witten, ``The $1/N$ expansion in atomic and particle physics,''
  in: {\it Recent developments in gauge theories,} ed. t' Hooft
  (Plenum Press 1980), p. 403.
   
\bibitem{Klebanov:2002ja}
  I.~R.~Klebanov, A.~M.~Polyakov,
  ``AdS dual of the critical O(N) vector model,''
  Phys.\ Lett.\  {\bf B550}, 213-219 (2002).
  [hep-th/0210114].
  
\bibitem{Aharony:2008ug}
  O.~Aharony, O.~Bergman, D.~L.~Jafferis, J.~Maldacena,
  ``N=6 superconformal Chern-Simons-matter theories, M2-branes and their gravity duals,''
  JHEP {\bf 0810}, 091 (2008).
  [arXiv:0806.1218 [hep-th]]. 
  
\bibitem{Denef:2008wq}
  F.~Denef,
  ``Les Houches Lectures on Constructing String Vacua,''
    [arXiv:0803.1194 [hep-th]].
  
\bibitem{Hartnoll:2009sz}
  S.~A.~Hartnoll,
  ``Lectures on holographic methods for condensed matter physics,''
  Class.\ Quant.\ Grav.\  {\bf 26}, 224002 (2009).
  [arXiv:0903.3246 [hep-th]]. 
  
  \bibitem{anderson}
P.~W.~Anderson, {\it Basic notions of condensed matter physics}, Addison-Wesley, 1984.

\bibitem{Polchinski:1992ed}
  J.~Polchinski,
  ``Effective Field Theory And The Fermi Surface,''
  arXiv:hep-th/9210046.
  
  \bibitem{shankar}
  R.~Shankar,
  ``Renormalization-group approach to interacting fermions,''
  Rev. Mod. Phys. {\bf 66}, 129 (1994).
  
  \bibitem{criticality}
  S.~Sachdev, B.~Keimer,
  ``Quantum criticality,''
  Physics Today {\bf 64}, 29 (2011), [arXiv:1102.4628 [cond-mat.str-el]].
  
\bibitem{review}
H.~v.~L\"ohneysen, A.~Rosh, M.~Vojta and P.~W\"olfle,
``Fermi-liquid instabilities at magnetic quantum phase transitions,''
Rev. Mod. Phys. {\bf 79}, 1015 (2007).
[arXiv:cond-mat/0606317 [cond-mat.str-el]].

\bibitem{subirbook}
 S.~Sachdev, {\it Quantum phase transitions}, CUP, 1999.

\bibitem{mottreview}
P.~A.~Lee, N.~Nagaosa and X.-G.~Wen,
``Doping a Mott insulator: Physics of high-temperature superconductivity,''
Rev. Mod. Phys. {\bf 78}, 17 (2006).
[arXiv:cond-mat/0410445 [cond-mat.str-el]].

\bibitem{Polyakov:1976fu}
  A.~M.~Polyakov,
  ``Quark Confinement and Topology of Gauge Groups,''
  Nucl.\ Phys.\  {\bf B120}, 429-458 (1977).

\bibitem{sungsik0}
S.-S.~Lee,
``Stability of the U(1) spin liquid with a spinon Fermi surface in 2+1 dimensions,''
Phys. Rev. {\bf B 78}, 085129 (2008).

\bibitem{Unsal:2008sc}
  M.~Unsal,
  ``Topological symmetry, spin liquids and CFT duals of Polyakov model with massless fermions,''
  [arXiv:0804.4664 [cond-mat.str-el]].

\bibitem{sungsik}
S.-S.~Lee,
``Low energy effective theory of Fermi surface coupled with a $U(1)$ gauge field in 2+1 dimensions,''
Phys. Rev. {\bf B 80}, 165102 (2009).
[arXiv:0905.4532 [cond-mat.str-el]].

\bibitem{Metlitski:2010pd}
  M.~A.~Metlitski and S.~Sachdev,
  ``Quantum phase transitions of metals in two spatial dimensions: I.
  Ising-nematic order,''
  arXiv:1001.1153 [cond-mat.str-el].

\bibitem{Metlitski:2010vm}
  M.~A.~Metlitski and S.~Sachdev,
  ``Quantum phase transitions of metals in two spatial dimensions: II. Spin
  density wave order,''
  Phys.\ Rev.\  B {\bf 82}, 075128 (2010)
  [arXiv:1005.1288 [cond-mat.str-el]].

\bibitem{mross} 
D.~F.~Mross, J.~McGreevy, H.~Liu and T.~Senthil, ``A controlled expansion for certain 
non-Fermi liquid metals,'' arXiv:1003.0894 [cond-mat.str-el].

\bibitem{Chamblin:1999tk}
  A.~Chamblin, R.~Emparan, C.~V.~Johnson, R.~C.~Myers,
  ``Charged AdS black holes and catastrophic holography,''
  Phys.\ Rev.\  {\bf D60}, 064018 (1999).
  [hep-th/9902170].

\bibitem{Hartnoll:2007ih}
  S.~A.~Hartnoll, P.~K.~Kovtun, M.~Muller, S.~Sachdev,
  ``Theory of the Nernst effect near quantum phase transitions in condensed matter, and in dyonic black holes,''
  Phys.\ Rev.\  {\bf B76}, 144502 (2007).
  [arXiv:0706.3215 [cond-mat.str-el]].

\bibitem{Hartnoll:2007ip}
  S.~A.~Hartnoll, C.~P.~Herzog,
  ``Ohm's Law at strong coupling: S duality and the cyclotron resonance,''
  Phys.\ Rev.\  {\bf D76}, 106012 (2007).
  [arXiv:0706.3228 [hep-th]].

\bibitem{Gibbons}
G.~W.~Gibbons,
``Aspects of supergravity theories,''
in: {\it Supersymmetry, Supergravity and Related Topics,} eds.
F.~del Aguila, J.~ de Azc\'arraga, and L. Ib\'a\~nez (World
Scientific, Singapore 1985), p. 147.

\bibitem{Jensen:2011su}
  K.~Jensen, S.~Kachru, A.~Karch, J.~Polchinski, E.~Silverstein,
  ``Towards a holographic marginal Fermi liquid,''
    [arXiv:1105.1772 [hep-th]].
  
\bibitem{Sen:2005wa}
  A.~Sen,
  ``Black hole entropy function and the attractor mechanism in higher derivative gravity,''
  JHEP {\bf 0509}, 038 (2005).
  [hep-th/0506177].
  
\bibitem{Hartnoll:2008vx}
  S.~A.~Hartnoll, C.~P.~Herzog, G.~T.~Horowitz,
  ``Building a Holographic Superconductor,''
  Phys.\ Rev.\ Lett.\  {\bf 101}, 031601 (2008).
  [arXiv:0803.3295 [hep-th]].
  
\bibitem{Hartnoll:2008kx}
  S.~A.~Hartnoll, C.~P.~Herzog, G.~T.~Horowitz,
  ``Holographic Superconductors,''
  JHEP {\bf 0812}, 015 (2008).
  [arXiv:0810.1563 [hep-th]].
   
\bibitem{Hartnoll:2009ns}
  S.~A.~Hartnoll, J.~Polchinski, E.~Silverstein, D.~Tong,
  ``Towards strange metallic holography,''
  JHEP {\bf 1004}, 120 (2010).
  [arXiv:0912.1061 [hep-th]].
  
\bibitem{Hartnoll:2010gu}
  S.~A.~Hartnoll, A.~Tavanfar,
  ``Electron stars for holographic metallic criticality,''
  Phys.\ Rev.\  {\bf D83}, 046003 (2011).
  [arXiv:1008.2828 [hep-th]].
  
\bibitem{Gubser:2008px}
  S.~S.~Gubser,
  ``Breaking an Abelian gauge symmetry near a black hole horizon,''
  Phys.\ Rev.\  {\bf D78}, 065034 (2008).
  [arXiv:0801.2977 [hep-th]].
  
\bibitem{Breitenlohner:1982bm}
  P.~Breitenlohner, D.~Z.~Freedman,
  ``Positive Energy in anti-De Sitter Backgrounds and Gauged Extended Supergravity,''
  Phys.\ Lett.\  {\bf B115}, 197 (1982).
  
\bibitem{Hertog:2006rr}
  T.~Hertog,
  ``Towards a Novel no-hair Theorem for Black Holes,''
  Phys.\ Rev.\  {\bf D74}, 084008 (2006).
  [gr-qc/0608075].
  
\bibitem{Gubser:2008pf}
  S.~S.~Gubser, A.~Nellore,
  ``Low-temperature behavior of the Abelian Higgs model in anti-de Sitter space,''
  JHEP {\bf 0904}, 008 (2009).
  [arXiv:0810.4554 [hep-th]].
  
\bibitem{Denef:2009tp}
  F.~Denef, S.~A.~Hartnoll,
  ``Landscape of superconducting membranes,''
  Phys.\ Rev.\  {\bf D79}, 126008 (2009).
  [arXiv:0901.1160 [hep-th]].
  
\bibitem{Klebanov:1999tb}
  I.~R.~Klebanov, E.~Witten,
  ``AdS / CFT correspondence and symmetry breaking,''
  Nucl.\ Phys.\  {\bf B556}, 89-114 (1999).
  [hep-th/9905104].
  
\bibitem{Pioline:2005pf}
  B.~Pioline, J.~Troost,
  ``Schwinger pair production in AdS(2),''
  JHEP {\bf 0503}, 043 (2005).
  [hep-th/0501169].
  
\bibitem{Herzog:2009xv}
  C.~P.~Herzog,
  ``Lectures on Holographic Superfluidity and Superconductivity,''
  J.\ Phys.\ A {\bf A42}, 343001 (2009).
  [arXiv:0904.1975 [hep-th]]. 
  
\bibitem{Horowitz:2010gk}
  G.~T.~Horowitz,
  ``Introduction to Holographic Superconductors,''
    [arXiv:1002.1722 [hep-th]].
    
\bibitem{Horowitz:2009ij}
  G.~T.~Horowitz, M.~M.~Roberts,
  ``Zero Temperature Limit of Holographic Superconductors,''
  JHEP {\bf 0911}, 015 (2009).
  [arXiv:0908.3677 [hep-th]].
    
\bibitem{Gubser:2009cg}
  S.~S.~Gubser, A.~Nellore,
  ``Ground states of holographic superconductors,''
  Phys.\ Rev.\  {\bf D80}, 105007 (2009).
  [arXiv:0908.1972 [hep-th]].
  
\bibitem{Gubser:2008wz}
  S.~S.~Gubser, F.~D.~Rocha,
  ``The gravity dual to a quantum critical point with spontaneous symmetry breaking,''
  Phys.\ Rev.\ Lett.\  {\bf 102}, 061601 (2009).
  [arXiv:0807.1737 [hep-th]].
  
\bibitem{Gauntlett:2009dn}
  J.~P.~Gauntlett, J.~Sonner, T.~Wiseman,
  ``Holographic superconductivity in M-Theory,''
  Phys.\ Rev.\ Lett.\  {\bf 103}, 151601 (2009).
  [arXiv:0907.3796 [hep-th]].
  
\bibitem{Gubser:2009gp}
  S.~S.~Gubser, S.~S.~Pufu, F.~D.~Rocha,
  ``Quantum critical superconductors in string theory and M-theory,''
  Phys.\ Lett.\  {\bf B683}, 201-204 (2010).
  [arXiv:0908.0011 [hep-th]].
  
\bibitem{Kachru:2008yh}
  S.~Kachru, X.~Liu, M.~Mulligan,
  ``Gravity Duals of Lifshitz-like Fixed Points,''
  Phys.\ Rev.\  {\bf D78}, 106005 (2008).
  [arXiv:0808.1725 [hep-th]].
  
\bibitem{Adams:2008zk}
  A.~Adams, A.~Maloney, A.~Sinha, S.~E.~Vazquez,
  ``1/N Effects in Non-Relativistic Gauge-Gravity Duality,''
  JHEP {\bf 0903}, 097 (2009).
  [arXiv:0812.0166 [hep-th]].
  
\bibitem{Copsey:2010ya}
  K.~Copsey, R.~Mann,
  ``Pathologies in Asymptotically Lifshitz Spacetimes,''
  JHEP {\bf 1103}, 039 (2011).
  [arXiv:1011.3502 [hep-th]].
  
\bibitem{Horowitz:1989bv}
  G.~T.~Horowitz, A.~R.~Steif,
  ``Space-Time Singularities in String Theory,''
  Phys.\ Rev.\ Lett.\  {\bf 64}, 260 (1990).
  
\bibitem{Anninos:2010sq}
  D.~Anninos, S.~A.~Hartnoll, N.~Iqbal,
  ``Holography and the Coleman-Mermin-Wagner theorem,''
  Phys.\ Rev.\  {\bf D82}, 066008 (2010).
  [arXiv:1005.1973 [hep-th]].
  
  \bibitem{son}
  D.~Nickel, D.~T.~Son,
  ``Deconstructing holographic liquids,''
    [arXiv:1009.3094 [hep-th]].
  
\bibitem{Faulkner:2009wj}
  T.~Faulkner, H.~Liu, J.~McGreevy, D.~Vegh,
  ``Emergent quantum criticality, Fermi surfaces, and AdS(2),''
    [arXiv:0907.2694 [hep-th]].  
    
\bibitem{Liu:2009dm}
  H.~Liu, J.~McGreevy, D.~Vegh,
  ``Non-Fermi liquids from holography,''
  Phys.\ Rev.\  {\bf D83 } (2011)  065029.
  [arXiv:0903.2477 [hep-th]].
  
\bibitem{Lee:2008xf}
  S.~-S.~Lee,
  ``A Non-Fermi Liquid from a Charged Black Hole: A Critical Fermi Ball,''
  Phys.\ Rev.\  {\bf D79}, 086006 (2009).
  [arXiv:0809.3402 [hep-th]].
  
\bibitem{Cubrovic:2009ye}
  M.~Cubrovic, J.~Zaanen, K.~Schalm,
  ``String Theory, Quantum Phase Transitions and the Emergent Fermi-Liquid,''
  Science {\bf 325}, 439-444 (2009).
  [arXiv:0904.1993 [hep-th]].
  
\bibitem{Cubrovic:2010bf}
  M.~Cubrovic, J.~Zaanen, K.~Schalm,
  ``Constructing the AdS dual of a Fermi liquid: AdS Black holes with Dirac hair,''
    [arXiv:1012.5681 [hep-th]].
    
\bibitem{Ruffini:1969qy}
  R.~Ruffini, S.~Bonazzola,
  ``Systems of selfgravitating particles in general relativity and the concept of an equation of state,''
  Phys.\ Rev.\  {\bf 187}, 1767-1783 (1969).

\bibitem{deBoer:2009wk}
  J.~de Boer, K.~Papadodimas, E.~Verlinde,
  ``Holographic Neutron Stars,''
  JHEP {\bf 1010}, 020 (2010).
  [arXiv:0907.2695 [hep-th]].

\bibitem{Arsiwalla:2010bt}
  X.~Arsiwalla, J.~de Boer, K.~Papadodimas, E.~Verlinde,
  ``Degenerate Stars and Gravitational Collapse in AdS/CFT,''
  JHEP {\bf 1101}, 144 (2011).
  [arXiv:1010.5784 [hep-th]].

\bibitem{Hartnoll:2011dm}
  S.~A.~Hartnoll, D.~M.~Hofman, D.~Vegh,
  ``Stellar spectroscopy: Fermions and holographic Lifshitz criticality,''
    [arXiv:1105.3197 [hep-th]].
    
\bibitem{Schutz:1970my}
  B.~F.~Schutz,
  ``Perfect Fluids in General Relativity: Velocity Potentials and a Variational Principle,''
  Phys.\ Rev.\  {\bf D2}, 2762-2773 (1970).
  
\bibitem{Hartnoll:2010ik}
  S.~A.~Hartnoll, P.~Petrov,
  ``Electron star birth: A continuous phase transition at nonzero density,''
  Phys.\ Rev.\ Lett.\  {\bf 106}, 121601 (2011).
  [arXiv:1011.6469 [hep-th]].
  
\bibitem{Puletti:2010de}
  V.~G.~M.~Puletti, S.~Nowling, L.~Thorlacius, T.~Zingg,
  ``Holographic metals at finite temperature,''
  JHEP {\bf 1101}, 117 (2011).
  [arXiv:1011.6261 [hep-th]].
  
\bibitem{Edalati:2011yv}
  M.~Edalati, K.~W.~Lo, P.~W.~Phillips,
  ``Neutral Order Parameters in Metallic Criticality in d=2+1 from a Hairy Electron Star,''
    [arXiv:1106.3139 [hep-th]].
  
\bibitem{Taylor:2008tg}
  M.~Taylor,
  ``Non-relativistic holography,''
    [arXiv:0812.0530 [hep-th]].

\bibitem{Goldstein:2009cv}
  K.~Goldstein, S.~Kachru, S.~Prakash, S.~P.~Trivedi,
  ``Holography of Charged Dilaton Black Holes,''
  JHEP {\bf 1008}, 078 (2010).
  [arXiv:0911.3586 [hep-th]].
  
\bibitem{Gubser:2009qt}
  S.~S.~Gubser, F.~D.~Rocha,
  ``Peculiar properties of a charged dilatonic black hole in $AdS_5$,''
  Phys.\ Rev.\  {\bf D81}, 046001 (2010).
  [arXiv:0911.2898 [hep-th]]. 
  
\bibitem{Charmousis:2010zz}
  C.~Charmousis, B.~Gouteraux, B.~S.~Kim, E.~Kiritsis, R.~Meyer,
  ``Effective Holographic Theories for low-temperature condensed matter systems,''
  JHEP {\bf 1011}, 151 (2010).
  [arXiv:1005.4690 [hep-th]].  
  
\bibitem{Iizuka:2011hg}
  N.~Iizuka, N.~Kundu, P.~Narayan, S.~P.~Trivedi,
  ``Holographic Fermi and Non-Fermi Liquids with Transitions in Dilaton Gravity,''
    [arXiv:1105.1162 [hep-th]].
  
\bibitem{Witten:1998zw}
  E.~Witten,
  ``Anti-de Sitter space, thermal phase transition, and confinement in gauge theories,''
  Adv.\ Theor.\ Math.\ Phys.\  {\bf 2}, 505-532 (1998).
  [hep-th/9803131].
  
\bibitem{ElShowk:2011ag}
  S.~El-Showk, K.~Papadodimas,
  ``Emergent Spacetime and Holographic CFTs,''
    [arXiv:1101.4163 [hep-th]].

\bibitem{subir}
S.~Sachdev, 
``Holographic metals and the fractionalized Fermi liquid,'' Phys. Rev. Lett. {\bf 105},
151602 (2010) [arXiv:1006.3794 [hep-th]].
  
\bibitem{Iqbal:2011in}
  N.~Iqbal, H.~Liu, M.~Mezei,
  ``Semi-local quantum liquids,''
    [arXiv:1105.4621 [hep-th]].
    
\bibitem{Huijse:2011hp}
  L.~Huijse, S.~Sachdev,
  ``Fermi surfaces and gauge-gravity duality,''
  [arXiv:1104.5022 [hep-th]].
  
   \bibitem{fraction} 
 T.~Senthil, S.~Sachdev, M.~Vojta, 
 ``Fractionalized Fermi liquids,"
 Phys. Rev. Lett. {\bf 90},
216403 (2003) [arXiv:cond-mat/0209144].
  
\end{thebibliography}
\end{document}